\documentclass{pasj00} 

\begin{document}
\SetRunningHead{T.~Inui \etal}{Time Variability of the Neutral Iron Lines from the Sgr B2 Region and its Implication of a Past Outburst of Sgr A$^*$}
\Received{2008/01/15}
\Accepted{2008/02/29}

\title{Time Variability of the Neutral Iron Lines from the Sgr B2 Region and its Implication of a Past Outburst of Sgr A$^*$}

\author{Tatsuya \textsc{Inui}, Katsuji \textsc{Koyama}, Hironori {\sc Matsumoto} and Takeshi Go {\sc Tsuru}}

\affil{Department of Physics, Graduate School of Science, Kyoto University, Sakyo-ku, Kyoto 606-8502}
\email{inuit@cr.scphys.kyoto-u.ac.jp}


\maketitle

\begin{abstract}
We investigate long-term X-ray behaviors from the Sgr B2 complex 
using archival data of the X-ray satellites Suzaku, XMM-Newton, 
Chandra and ASCA.
The observed region of the Sgr B2 complex includes two prominent spots 
in the Fe\emissiontype{I} K-$\alpha$ line at 6.40~keV, 
a giant molecular cloud M\,0.66$-$0.02 known as the ``Sgr B2 cloud'' and 
an unusual X-ray source G\,0.570$-$0.018.
Although these 6.40~keV spots have spatial extensions of a few pc scale,
the morphology and flux of the 6.40~keV line has been time variable for 10 years, in contrast to the constant flux of the
Fe\emissiontype{XXV}-K$\alpha$ line at 6.67~keV 
in the Galactic diffuse X-ray emission.  
This time variation is mostly due to M\,0.66$-$0.02; 
the 6.40~keV line flux declined in 2001 and decreased to 60\% 
in the time span 1994--2005.
The other spot G\,0.570$-$0.018 is found to be conspicuous 
only in the Chandra observation in 2000.
From the long-term time variability ($\sim$10 years) 
of the Sgr B2 complex, we infer that 
the Galactic Center black hole Sgr A$^*$ was X-ray bright 
in the past 300 year and exhibited a time variability with a period of a few years. 
\end{abstract}

\section{Introduction}

Our Galactic Center (GC) black hole, Sgr~A$^*$, has a mass of 
$4\times 10^{6}M_\odot$ (\cite{Eisenhauer2005}; \cite{Ghez2005}), 
but is quiescent with an X-ray luminosity of only 
$10^{33\mbox{--}34}$ ergs~s$^{-1}$ \citep{Baganoff2001}, 
which is many orders of magnitude smaller than that of canonical AGNs.  
Sgr~A$^*$ has been also found to exhibit frequent X-ray flares 
(e.g., \cite{Baganoff2001}).
This X-ray flare activity is not huge, however,
with a typical duration of 10~ks and an amplitude 
that is only a factor of ten above the non-flare phase.
Therefore, even at the flare peak, the X-ray luminosity is 
$\sim10^{6}$ times lower than the Eddington luminosity 
of this supermassive black hole.

The 6.40~keV line features in molecular clouds suggest an active phase in the past.
The Sgr B complex of giant molecular clouds emit peculiar X-rays 
with a strong K-shell transition line at 6.40~keV from neutral iron 
(equivalent width of 1--2~keV) and a deep absorption edge at 7.1~keV 
(equivalent $N_{\rm H}$ of $10^{23\mbox{--}24}$) \citep{Koyama1996}.
The 6.40~keV line flux and the spectrum necessitate the existence of strong X-ray sources irradiating the clouds.
The detailed X-ray morphology of the neutral iron line 
in the Sgr B2 cloud has been investigated by Chandra and it was found
that the external source is likely to be located 
in the GC direction \citep{Murakami2001sgrb}.
However, no such source has been found, neither toward the GC, 
nor in regions near the GC.
Based on these results, \citet{Koyama1996}, \citet{Murakami2001sgrb} 
and \citet{Koyama2007b} proposed that the Sgr B2 complex contains 
X-ray reflection nebulae (XRNe) irradiated by Sgr A$^*$, 
which must have been six orders of magnitude X-ray brighter 
about 300 years ago, which is the light travel time 
between Sgr B2 and Sgr A$^*$.
Other molecular clouds such as 
M\,0.74$-$0.09, Sgr B1, Sgr C and M\,0.11$-$0.08 also have intense 
6.40~keV lines (\cite{Koyama2007b}, \cite{Nobukawa2008}, 
\cite{Murakami2001sgrc}, \cite{Yusef2007}).
Some, if not all, of them, are also likely to be XRNe 
(\cite{Koyama2007b}, \cite{Nobukawa2008}, \cite{Murakami2001sgrc}). 

Since the 6.40~keV line can be also produced by 
high-energy electrons bombarding the cloud \citep{Tatischeff2003}, 
\citet{Predehl2003} and \citet{Yusef2007} argued against 
the XRN scenario and proposed electron bremsstrahlung instead.
One way to distinguish between these two possible scenarios (i.e., irradiation by X-rays or electrons (charged-particles)) would be to detect the 
time variability of the 6.40~keV line.   

\citet{Muno2007} discovered the time variability of 
the 6.40~keV line clumps in the molecular cloud M\,0.11$-$0.11.
\citet{Koyama2008a} also found changes in the flux and morphology of
the 6.40~keV line of the Sgr B2 complex between the Chandra and Suzaku 
observations in 2000 and 2005, respectively.
   
Since variability over a few years from a pc-scale object 
requires light-speed communication, 
charged particles are unlikely to be the origin of the 6.40~keV line.
This time variability, on the other hand, supports the XRN scenario; 
X-rays from external sources irradiate molecular clouds and produce 
the fluorescence iron line at 6.40~keV and 
Thompson scattering continuum emission in the same manner as an X-ray echo.
Using the time variable X-ray echo, 
we can infer the light curve of Sgr A$^*$ in the past. 

This paper reports a comprehensive X-ray study based on 
all the available archival data of the Sgr B2 complex 
(i.e., the ASCA, Chandra, XMM-Newton and Suzaku data).
We derived a 10-year-scale light curve for the 6.40~keV line. 
We found the time variability of the 6.40~keV line from the Sgr B2 complex by using the time-constant 6.67~keV line from the GC diffuse X-rays (GCDX) as a reference.
In this paper, the distance to the Sgr B2 complex 
is assumed to be 8.5~kpc \citep{Reid1988}.

\section{Observations and Data Reduction}

We used four deep exposure and two short survey observations 
as listed in table \ref{tab:obs_data}. 
However, for imaging analysis
we used only the deep exposure observations.

\begin{table*}
  \caption{Observation Log}
  \label{tab:obs_data}
  \begin{center}
    \begin{tabular}{lrcr}
      \hline\hline
      Observatory & Obs. ID & Date & Exposure Time \\
       & & (yyyy/mm/dd) & (sec) \\
      \hline
      ASCA & 52006000 & 1994/09/22 & 58,395 \\
      ASCA & 52006001 & 1994/09/24 & 21,500 \\
      Chandra & 944 & 2000/03/29 & 98,629 \\
      XMM-Newton & 0112971501 & 2001/04/01 & 12,649 \\
      Chandra & 2280 & 2001/07/16 & 10,423 \\
      XMM-Newton & 0203930101 & 2004/09/04 & 42,282 \\
      Suzaku & 100037060 & 2005/10/10 & 89,072 \\
      \hline
    \end{tabular}
  \end{center}
\end{table*}

\subsection{Suzaku}

The Sgr B2 complex was observed with the XIS on 10--12 October 2005.
The XIS consists of four sets of X-ray CCD camera systems 
(XIS0, 1, 2, and 3) placed on the focal planes of four X-ray telescopes
(XRTs) on board the Suzaku satellite.
XIS0, 2 and 3 have front-illuminated (FI) CCDs, 
whereas XIS1 has a back-illuminated (BI) CCD. 
Detailed descriptions of the Suzaku satellite, 
XRT and XIS are found in 
\citet{Mitsuda2007}, \citet{Serlemitosos2007} and \citet{Koyama2007a}.
The XIS observation was made in the normal mode. 
The effective exposure time after removing the epoch of 
the low earth elevation angle (ELV $\le$5$^\circ$ ) 
and the South Atlantic Anomaly was about 89~ks.
We analyzed the data using the software package HEASoft 6.2.
The XIS gain was fine-tuned using the same procedures 
described in \citet{Koyama2007b}.
Using point sources in the field, 
we re-registered the absolute astrometry
by shifting $(-\timeform{0D.0001},~ -\timeform{0D.0062})$ 
in the $(l,~ b)$ coordinate,
referring the Chandra point-source positions (\cite{Koyama2007b}).

\subsection{XMM-Newton}

The X-ray data were obtained using the European Photon Imaging Camera 
(EPIC) (\cite{Struder2001}; \cite{Turner2001}) 
onboard the XMM-Newton on 1 April 2001 and 4 September 2004.
The observations were performed in extended full-frame mode. 
The data were analyzed using Science Analysis Software (SAS 7.1.0).
Event files for both the PN and the Metal Oxide Semiconductor (MOS) 
detectors were produced 
using the epproc and emproc tasks of SAS, respectively.
The event files were screened for high particle-background periods.
In our analysis, we dealt only with events corresponding to 
patterns 0--4 for the PN and 0--12 for the MOS instruments.
For the observation in 2004, we made the total band images 
(0.3--10~keV) for each detector and detected point sources 
(93 point sources from MOS1, 97 from MOS2, and 111 from PN).
27 sources were detected with all three detectors 
within positional uncertainties of 10~arcsec, 
the sizes of PSF of EPIC at 6~keV, 
and 19 sources have possible counterparts 
with catalogued point sources of Chandra \citep{Muno2003}.
The mean position difference of these point sources 
relative to the Chandra positions is 
$(-\timeform{0D.0006},~ \timeform{0D.0003})$ 
in celestial coordinates.
We fine-tuned the XMM-Newton frame by shifting 
$(\Delta\alpha,~\Delta\delta)
=(\timeform{0D.0006},~-\timeform{0D.0003})$.

\subsection{Chandra}

Sgr B2 was observed using the Advanced CCD Imaging Spectrometer 
(ACIS-I) onboard the Chandra observatory \citep{Weisskopf2002} 
on 29 March 2000 and 16 July 2001.
We used the event files provided by standard pipeline processing.
Only the grade 0, 2, 3, 4, and 6 events were used in the analysis.
Each CCD chip subtends an 8.3-arcmin square in the sky, 
while the pixel size is 0.5 arcsec.
The on-axis spatial resolution is 0.5-arcsec 
(full width at half-maximum (FWHM)).
Images, spectra, ancillary files and response matrices 
were created using the software CIAO v3.4.
The absolute positional accuracy is 0.6~arcsec \citep{Weisskopf2003}.

\subsection{ASCA}

The ASCA observations of Sgr B2 were performed on 22 and 24 September 1994 for $\sim 80 \mathrm{ks}$ in total.
We used the event files provided by standard pipeline processing.
We merged the two observation data since the observation dates and aim points are almost the same.
Since the GIS data did not have sufficient energy resolution 
($\sim 470$~eV at 5.9~keV) to separate the iron lines, 
we used only SIS data for the analysis.
Unlike Suzaku and XMM-Newton, no available point sources 
to use as a reference for Chandra for the absolute astrometry correction 
were found.
Therefore, the astrometric uncertainty of SIS image 
remains about 40~arcsec\footnote{See http://heasarc.gsfc.nasa.gov/docs/asca/cal\_probs.html}.
We used the software package HEASoft 6.2 in the analysis.

\section{Analysis}
\subsection{Spectra of the Sgr B2 region}
\begin{figure}
  \begin{center}
    \FigureFile(80mm,50mm){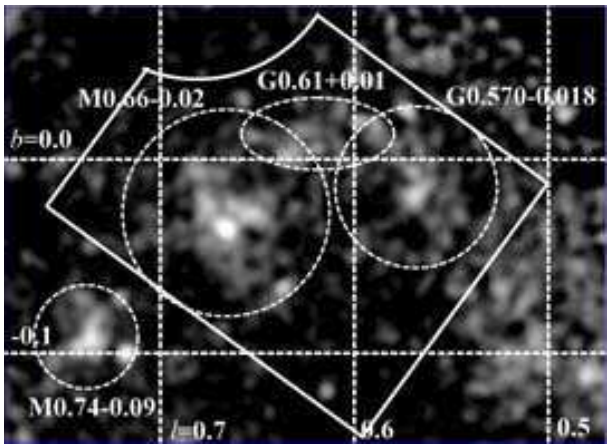}
  \end{center}
  \caption{XMM-Newton image in the 6--7~keV band. 
    The solid line indicates the selected area of the Sgr B2 complex (see text). 
    The dashed circles are the individual diffuse sources 
    discussed in this paper. 
    M\,0.66$-$0.02 is the Sgr B2 cloud. 
    G\,0.61+0.01 is a new source with a strong 6.67~keV line, 
    possibly a new young SNR.     G\,0.570$-$0.0018 is an unusual diffuse source. 
}
\label{img_sgrb_overall_xmm}
\end{figure}

Figure \ref{img_sgrb_overall_xmm} shows relevant diffuse sources 
in the Sgr B2 complex overlaid on the XMM-Newton 6--7~keV image.
Since the FOVs and pointing positions of 
the Suzaku, Chandra, XMM-Newton and ASCA observations 
are slightly different from each other, 
we selected a region in which all the observed areas overlapped 
(referred to hereafter as the Sgr B2 region).
The Sgr B2 region (the overlapping region) is indicated by 
by the solid squares in figure \ref{img_sgrb_overall_xmm}.
The subsequent data analysis and discussion are based on 
the sources in the Sgr B2 region.
Hereafter, we refer to the Sgr B2 cloud as M\,0.66$-$0.02. G\,0.61+0.01 
is a new source with a strong 6.67~keV line, 
hence possibly a new young SNR \citep{Koyama2007b}. 
G\,0.570$-$0.0018 is an unusual diffuse source discovered 
by \cite{Senda2002}. 
M\,0.74$-$0.09 is also a new 6.40~keV source, 
but it is not discussed in this paper, 
because it is outside of the Sgr B2 region.   

We extracted X-ray spectra from the Sgr B2 region 
and subtracted the off-plane blank-sky regions.
Thus, the cosmic X-ray background (CXB) 
and non-X-ray background (NXB) were subtracted 
but the GCDX were not subtracted.
As the off-plane blank-sky, we selected 
the north ecliptic pole data for XIS, 
distributed blank-sky databases for MOS and ACIS, 
and Lockman hole data for PN and SIS.

\begin{figure*}
  \begin{center}
    \FigureFile(50mm,50mm){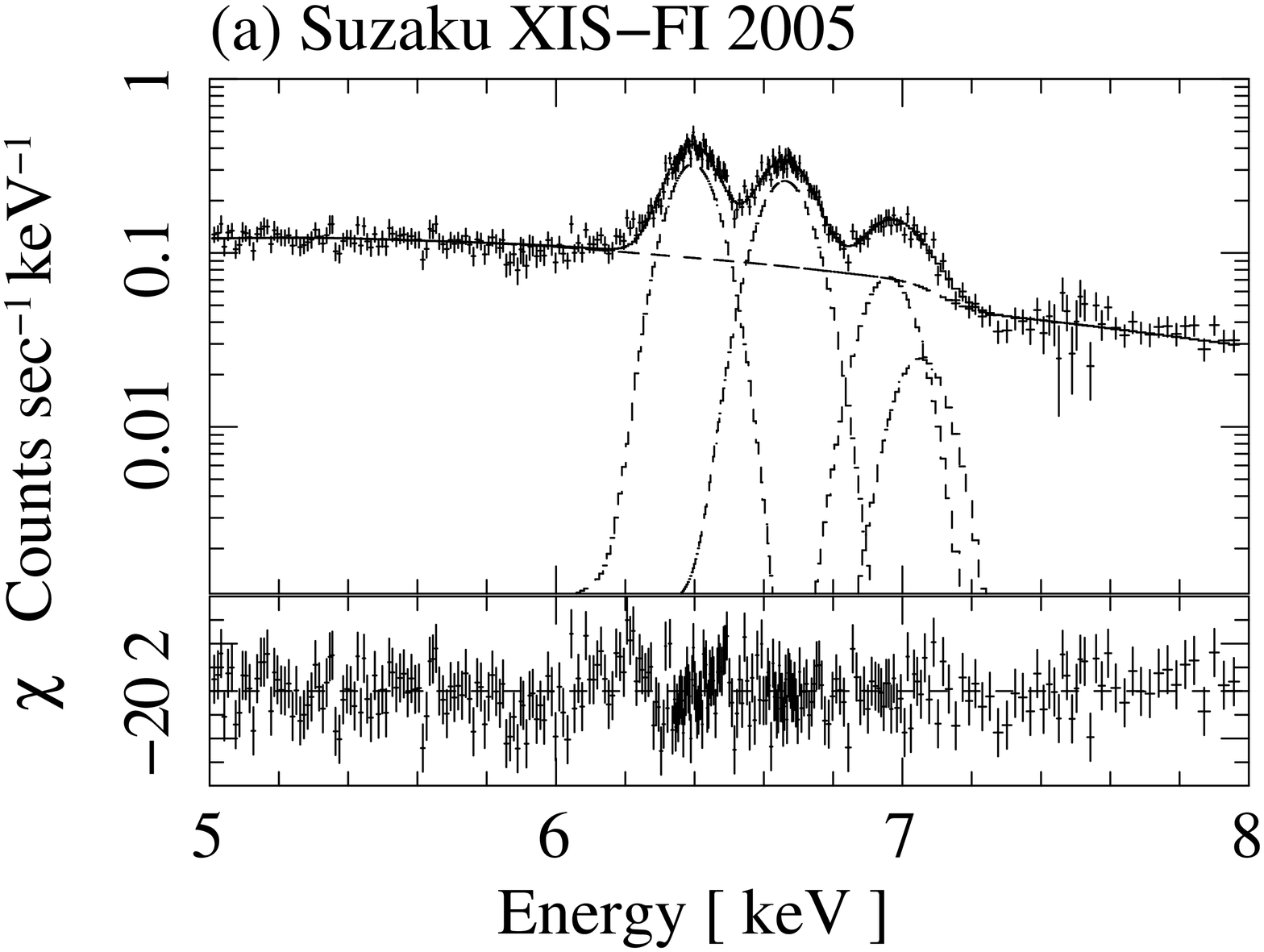}
    \FigureFile(50mm,50mm){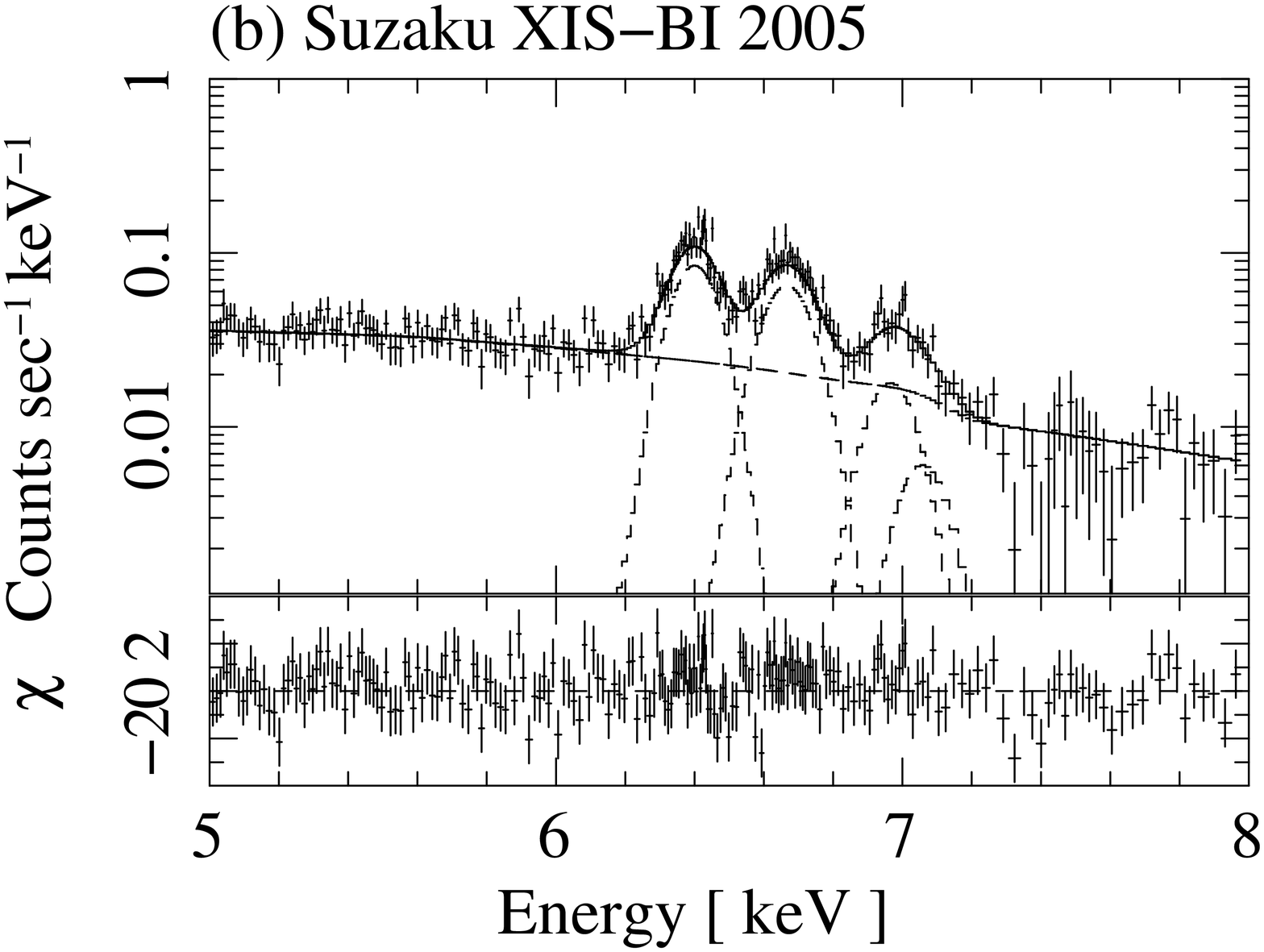}
    \FigureFile(50mm,50mm){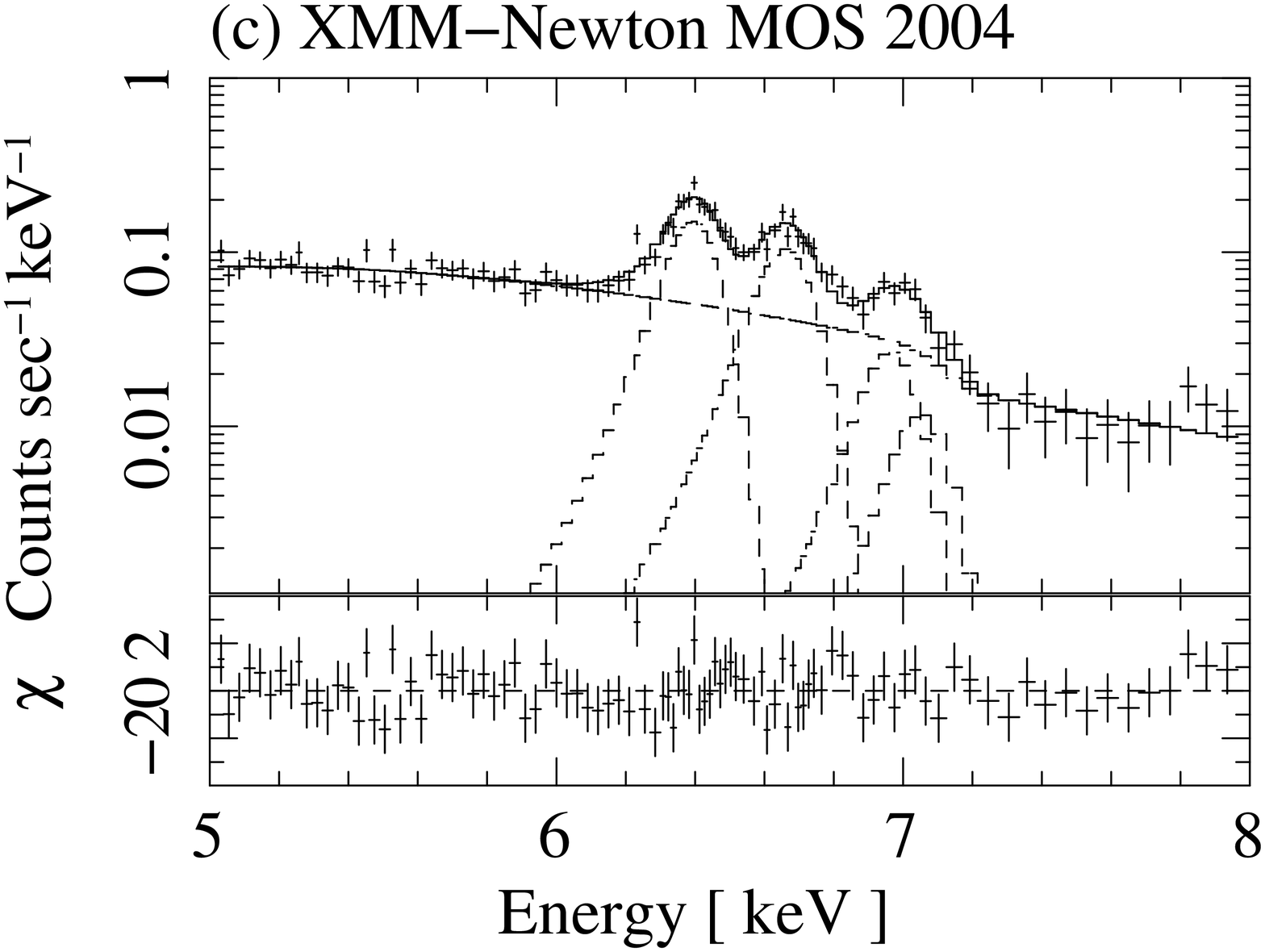}
    \FigureFile(50mm,50mm){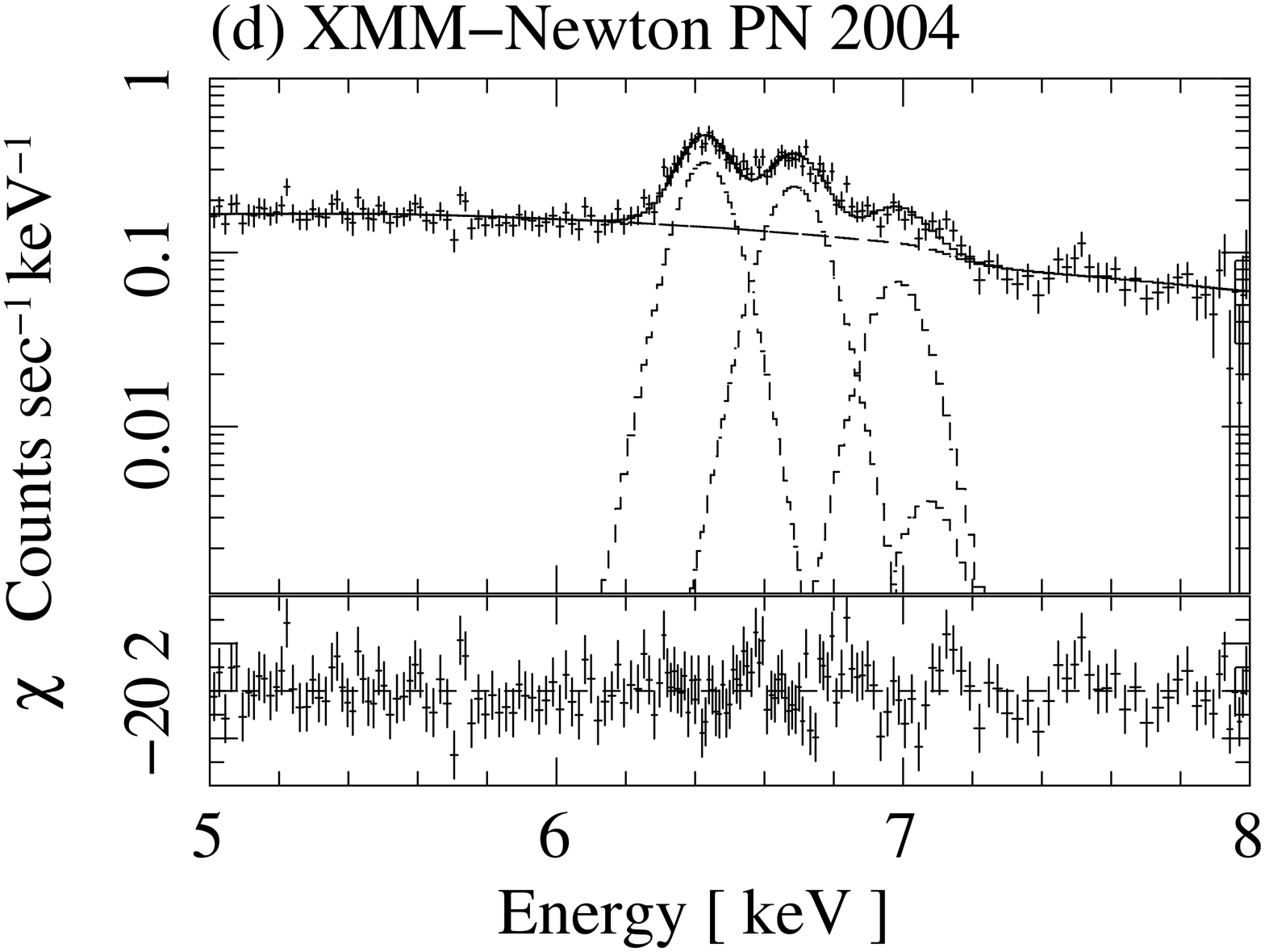}
    \FigureFile(50mm,50mm){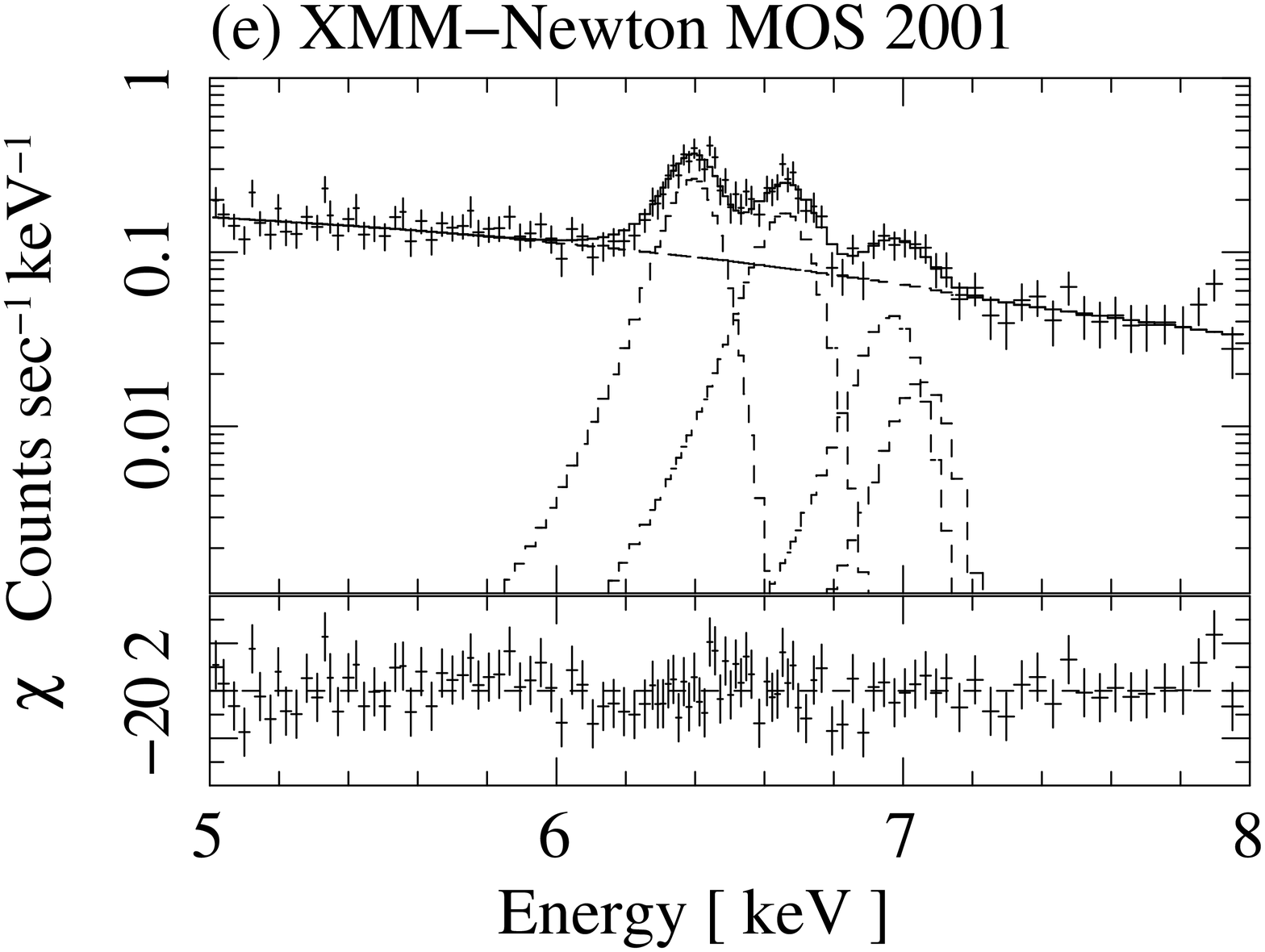}
    \FigureFile(50mm,50mm){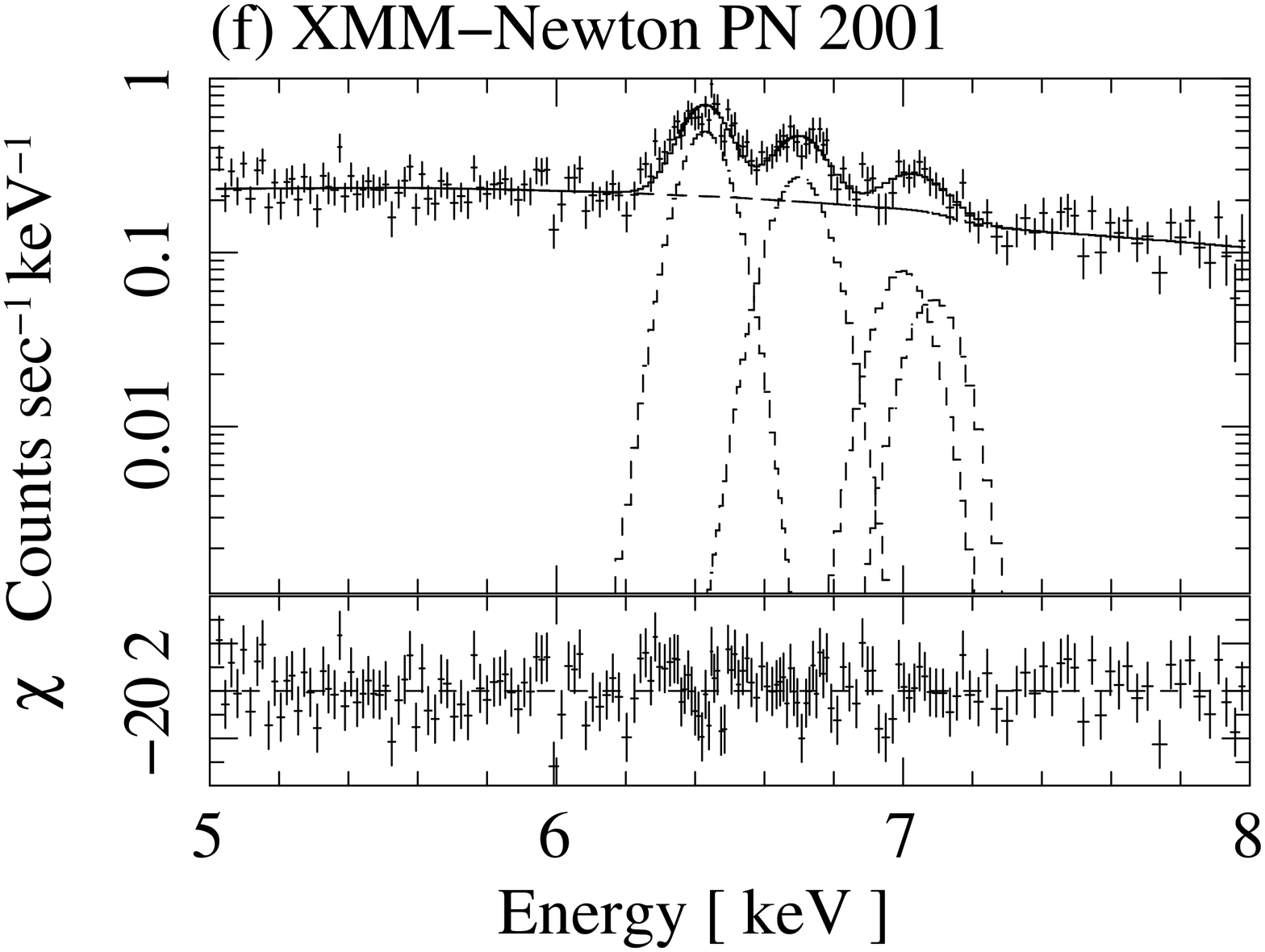}
    \FigureFile(50mm,50mm){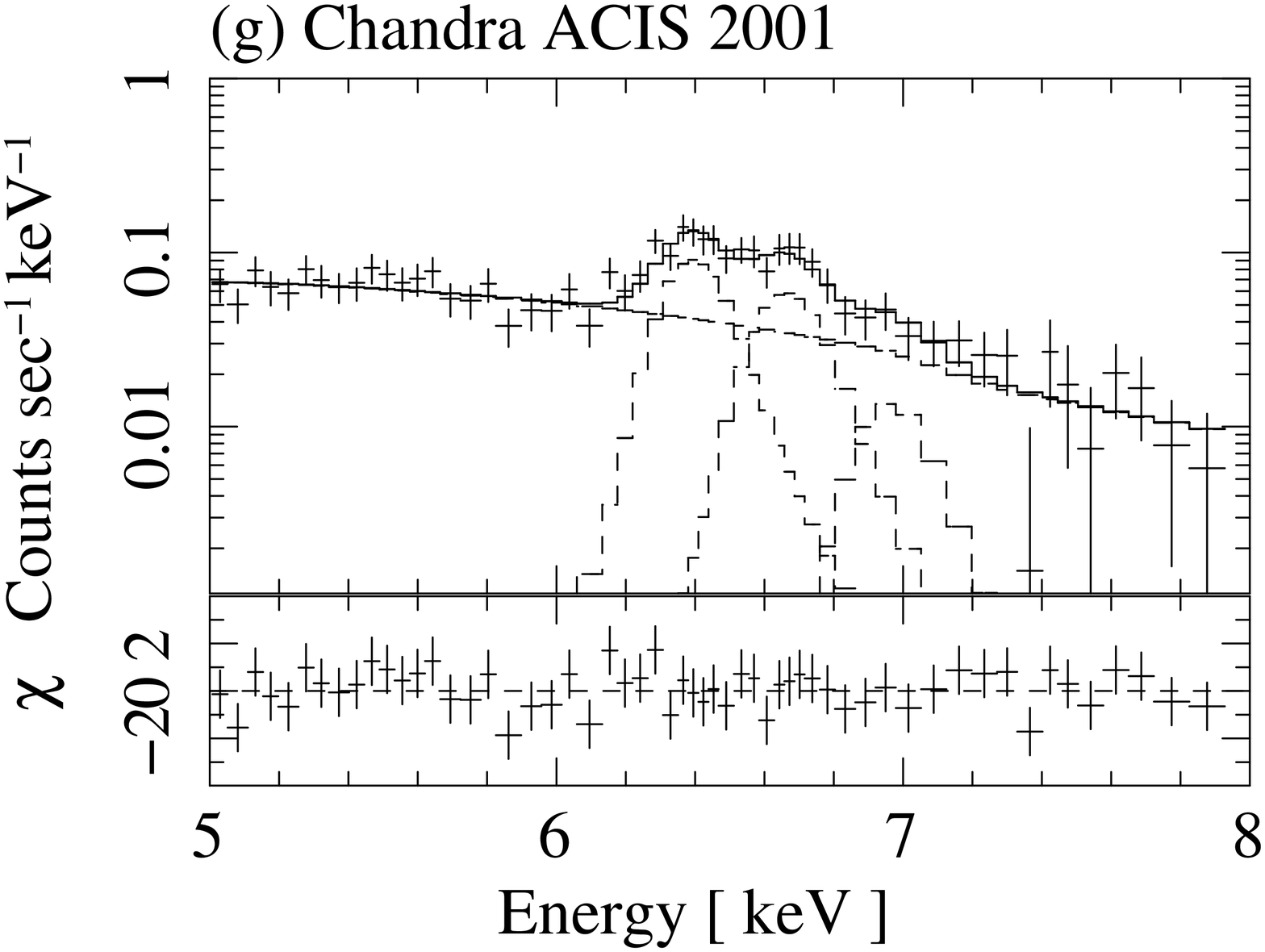}
    \FigureFile(50mm,50mm){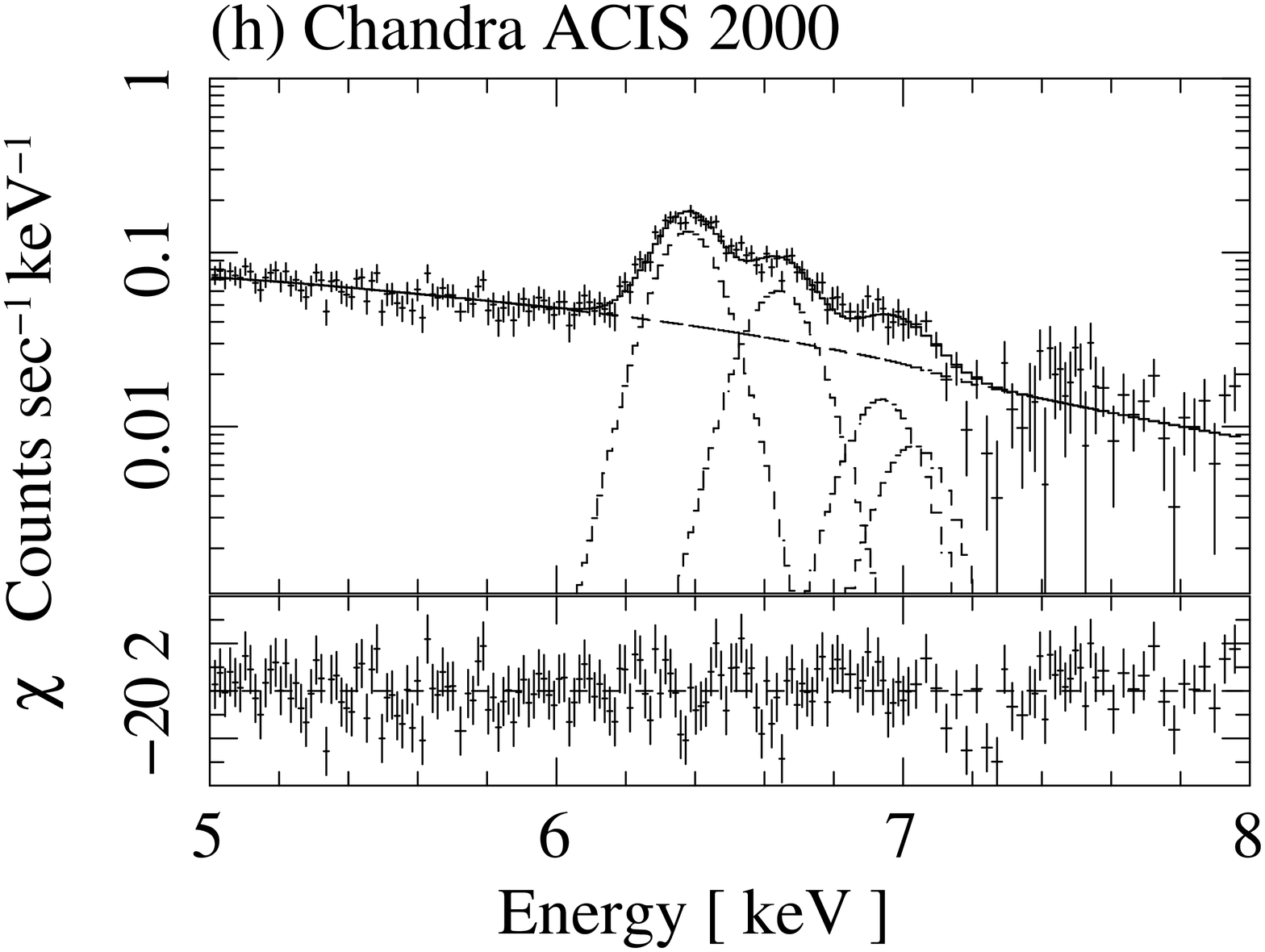}
    \FigureFile(50mm,50mm){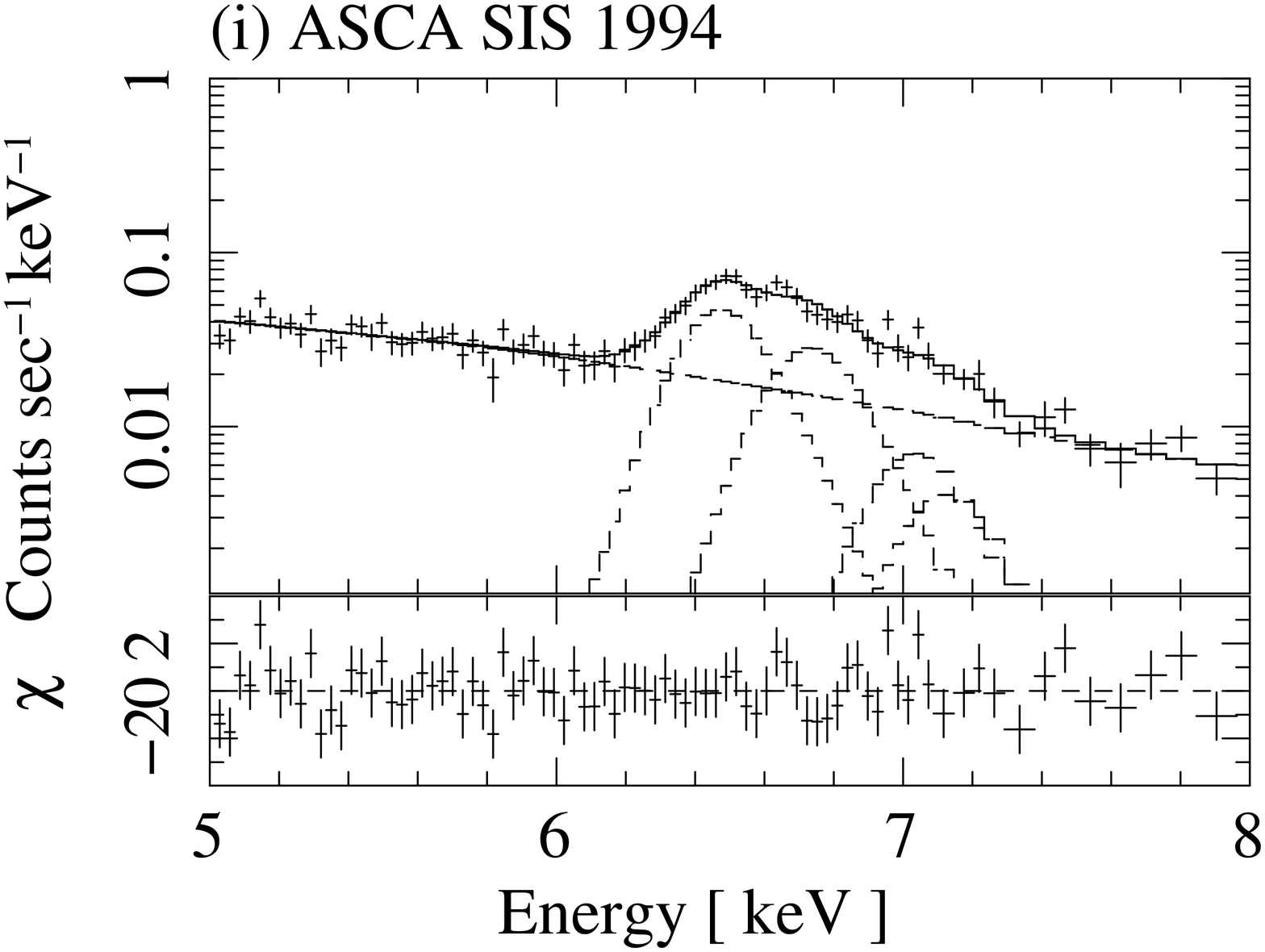}
  \end{center}
  \caption{The blank-sky-subtracted spectra of the Sgr B2 region, 
    with the Suzaku XIS FI sensors (a) and BI sensor (b) in 2005,
    XMM-Newton MOS (c) and PN (d) in 2004, (e) and (f) in 2001, 
    Chandra ACIS-I (g) in 2001 and (h) in 2000, 
    and ASCA SIS (i) in 1994. 
    The solid lines indicate the best-fit models (see text).}
  \label{fig:spec_b2area}
\end{figure*}

The blank-sky-subtracted spectra are shown 
in figure \ref{fig:spec_b2area}. 
These spectra exhibit continuum emissions 
and pronounced peaks due to Fe \emissiontype{I} K$\alpha$ (6.40~keV), 
Fe \emissiontype{XXV} K$\alpha$ (6.67~keV), 
and the composite of Fe \emissiontype{XXVI} K$\alpha$ (6.96~keV) 
and Fe \emissiontype{I} K$\beta$(7.06~keV).
The continuum emission is composed of: 
1) the continuum of the GCDX (\cite{Koyama2007c}), 
2) the continuum (thermal bremsstrahlung) from G\,0.61+0.01, 
a new SNR in this region (\cite{Koyama2007b}), 
and 3) a non-thermal power-law continuum with a deep iron K-edge 
related to the 6.40~keV line (\cite{Koyama1996}).

It is practically impossible to resolve these three continuums; 
we therefore represented these three continuum 
by a single power-law continuum with absorption 
(the wabs*power-law model in the xspec code).
We then added four K-shell lines by Gaussian functions.
These are the Fe \emissiontype{XXV} K$\alpha$ ($\sim$6.67~keV) 
and Fe \emissiontype{XXVI} K$\alpha$ ($\sim$6.96~keV) lines 
due to the GCDX, and Fe \emissiontype{I} K$\alpha$ (6.40~keV) 
and Fe \emissiontype{I} K$\beta$(7.06~keV) lines 
due mainly to the Sgr B2 region.
Thus, the model spectrum in the xspec code is given as;

\begin{equation}
\mbox{wabs*power-law + 4 Gaussians}
\end{equation}

\subsubsection{Suzaku}

The spectra of the Sgr B2 region and background subtractions 
of the blank sky for XIS 0,1,2 and 3 were performed separately, 
however all the FI-CCD spectra (XIS0, 1 and 3) were combined, 
because their response functions are almost identical with each other.
We fitted the FI-CCD and BI-CCD spectra simultaneously 
in the energy band of 5--8 keV with the model given by equation (1). 
All the parameters were free under the following constraints: 
the line widths were fixed to be 0~eV except 
that of the Fe \emissiontype{XXV} K$\alpha$ line \citep{Koyama2007c}, 
and the energy interval between Fe \emissiontype{I} K$\alpha$ (6400~eV)
and K$\beta$ (7058~eV) was fixed at the theoretical value of 658~eV 
\citep{Kaastra1993}. 
The best-fit spectra for the FI and BI are shown in figures 2(a) and 2(b) respectively, 
while the best-fit parameters are listed 
in table \ref{tab:specfit_b2area_szk}.

Although the K-shell lines from highly ionized iron, 
the Fe \emissiontype{XXVI} K$\alpha$ 
and Fe \emissiontype{XXV} K$\alpha$ lines are due to 
the largely extended GCDX, the best-fit center energy of 
the Fe \emissiontype{XXVI} K$\alpha$ line (6665 eV) 
and the flux ratio of Fe \emissiontype{XXVI} K$\alpha$ 
to Fe \emissiontype{XXV} K$\alpha$ (0.30) are slightly smaller than 
those (6680 eV and 0.34, respectively) found in the GC region 
\citep{Koyama2007c}.
This may be due to the contamination of 
a strong Fe \emissiontype{XXV} K$\alpha$ line of a new SNR G\,0.61+0.01
\citep{Koyama2007b}. 

When fitting the XMM-Newton, Chandra and ASCA spectra, 
we fixed the line centroids and the K$\alpha$ flux ratio of 
Fe \emissiontype{XXVI} and Fe \emissiontype{XXV} 
to the best-fit Suzaku values 
given in table \ref{tab:specfit_b2area_szk}, 
because the statistics and energy response of these satellites 
were limited compared to Suzaku.

\begin{table}
  \caption{The best-fit Suzaku (XIS) results of the Sgr B2 region}
  \label{tab:specfit_b2area_szk}
  \begin{center}
    \begin{tabular}{llc}
      \hline\hline
       Parameter & Terminology & Value \\
      \hline
\multicolumn{3}{c}{---Continuum---}\\
Absorption\footnotemark[$a$] & 	$N_{\rm H}$ & 	$3.0_{-0.5}^{+0.3}$ \\
Photon index	& $\Gamma$ & 	$2.6_{-0.2}^{+0.2}$ \\
Power-law flux\footnotemark[$b$]& $F_{\rm pow}$ & $5.2_{-0.4}^{+0.4}$ \\
\multicolumn{3}{c}{---Neutral iron lines---}\\
Center energy of Fe\emissiontype{I}-K$\alpha$\footnotemark[$c$]&$E_{640}$ & $6400_{-1}^{+4}$\\
Flux of Fe \emissiontype{I}-K$\alpha$\footnotemark[$d$]&$F_{640}$ & $1.14_{-0.03}^{+0.04}$\\
Center energy of Fe \emissiontype{I}-K$\beta$\footnotemark[$c$]&$E_{706}$ & 7058\footnotemark[$f$]\\ 
Flux of Fe \emissiontype{I}-K$\beta$\footnotemark[$d$] &$F_{706}$ & $0.11_{-0.07}^{+0.05}$\\
      \multicolumn{3}{c}{---Highly ionized iron lines---}\\
Center energy of Fe \emissiontype{XXV}-K$\alpha$\footnotemark[$c$]&$E_{667}$ & $6665_{-3}^{+4}$ \\
Width of Fe \emissiontype{XXV}-K$\alpha$\footnotemark[$e$]&$\sigma_{667}$ & $27_{-9}^{+7}$ \\
Flux of Fe \emissiontype{XXV}-K$\alpha$\footnotemark[$d$]&$F_{667}$ & $1.07_{-0.05}^{+0.04}$ \\
Center energy of Fe \emissiontype{XXVI}-K$\alpha$\footnotemark[$c$]&$E_{697}$ & $6966_{-13}^{+14}$ \\
Flux of Fe \emissiontype{XXVI}-K$\alpha$\footnotemark[$d$]&$F_{697}$ & $0.32_{-0.04}^{+0.05}$\\
\hline
& $\chi^2$/dof & 574.4/514\\
\hline
\multicolumn{3}{@{}l@{}}{\hbox to 0pt{\parbox{85mm}{\footnotesize
Note---Parentheses indicate the 90\% confidence limit.
\par\noindent
\footnotemark[$a$] In unit of  $10^{23}$~H~cm$^{-2}$.
\par\noindent
\footnotemark[$b$]
Observed flux of a power-law continuum in the 5--8~keV band in unit of $10^{-4}$ photons cm$^{-2}$ s$^{-1}$.
\par\noindent
\footnotemark[$c$]
Line center energy in unit of eV.
\par\noindent
\footnotemark[$d$]
Observed line flux in unit of $10^{-4}$ photons cm$^{-2}$ s$^{-1}$.
\par\noindent
\footnotemark[$e$]
Line width in the Gaussian sigma in unit of eV. 
\par\noindent
\footnotemark[$f$] The energy gap between K$\alpha$ (6400~eV) and K$\beta$ (7058~eV) is fixed 
at the theoretical value (+658~eV) \citep{Kaastra1993}.
}\hss}}
  \end{tabular}
  \end{center}
\end{table}

\subsubsection{XMM-Newton}

We fitted the MOS and PN spectra separately, 
where the line center energies and the K$\alpha$ line flux ratio 
of Fe\emissiontype{XXVI} and Fe\emissiontype{XXV} were fixed to 
those of the Suzaku best-fit values (see Section 3.1.1). 
This fit failed in reproducing the line and edge structures 
in the PN spectra, indicating a systematic gain shift; 
we thus added a gain offset as a free parameter. 
The fit was then acceptable as 
Table \ref{tab:specfit_b2area_xmmcxoasca} shows. 
The best-fit spectra are given in figures 2(c)--(f).

\subsubsection{Chandra}

We fitted the Chandra spectra using the same method as the XMM-Newton case. 
The spectrum in 2000 was well fitted 
by adding a gain offset of $\sim -20$~eV.
The best-fit spectra and parameters are given 
in figure 2(g) and table \ref{tab:specfit_b2area_xmmcxoasca}. 
In the 2001 spectra, however, 
the best-fit photon index was unrealistic with $\Gamma\sim -2.2$.
If we fixed $\Gamma=2.6$, the best-fit Suzaku value, 
then there were large residuals above the 7~keV band.
We compared the count rates in the 10.0--12.5~keV band 
between the source and blank-sky spectra, 
and found that the high-energy band count-rate of the blank-sky 
was 2.4 times smaller than that of the source spectra.
Thus, the strange behaviors are likely due to unstable NXB.
We therefore subtracted the blank-sky spectra 
by first multiplying them by a factor of 2.4. 
We then obtained a good fit with reasonable parameters.
The best-fit spectra and parameters are shown 
in figure 2(h) and table \ref{tab:specfit_b2area_xmmcxoasca}, respectively.
It could be argued that the background subtraction described above is artificial, 
and hence may cause large systematic errors in the line flux.
However, the line flux errors for the different background subtractions 
are within 10\%.
More details concerning the line flux dependence on the continuum shape 
and/or background subtractions are given in Section 3.2.2, 
for the case of the XMM-Newton 
(table \ref{tab:specfit_b2area_xmm_multinxb}).

\subsubsection{ASCA}

The fitting procedures for the ASCA spectra were 
the same as those for XMM-Newton and Chandra spectra.
The ASCA spectra require a large systematic gain offset about 80~eV 
to adjust the iron line centroids.
This large offset is caused by chip-to-chip variation of the gain\footnote{See http://heasarc.gsfc.nasa.gov/docs/asca/4ccd.html}.
The best-fit spectra and parameters are shown 
in figure 2(i) and in table \ref{tab:specfit_b2area_xmmcxoasca}.

\begin{table*}
  \caption{The XMM-Newton, Chandra and ASCA fitting results for the Sgr B2 region}
  \label{tab:specfit_b2area_xmmcxoasca}
  \begin{center}
    \begin{tabular}{lccccccc}
      \hline
  Observatory & \multicolumn{2}{c}{XMM-Newton} & \multicolumn{2}{c}{XMM-Newton} & \multicolumn{2}{c}{Chandra} & ASCA \\
  Detector & MOS(2004) & PN(2004) & MOS(2001) & PN(2001) & ACIS(2001) & ACIS(2000) &SIS(1994) \\
      \hline
	\multicolumn{8}{c}{---Continuum ---}\\
$N_{\rm H}$& $5.3_{-1.3}^{+1.4}$ &$2.1_{-0.4}^{+0.4}$ &	$0.5_{-0.5}^{+1.3}$ & $1.5_{-0.9}^{+1.0}$ &$2.4_{-2.4}^{+3.4}$ & $1.0_{-0.4}^{+0.5}$ & $0.7_{-0.7}^{+1.0}$ \\
$\Gamma$& $3.6_{-0.7}^{+0.7}$ &	$1.6_{-0.1}^{+0.2}$ & $0.4_{-0.5}^{+0.6}$ & $0.9_{-0.4}^{+0.4}$ &$1.0_{-1.5}^{+1.7}$ & $0.7_{-0.5}^{+0.4}$& $1.1_{-0.5}^{+0.7}$ \\
$F_{\rm pow}$ & $4.8_{-0.2}^{+0.2}$ &$7.0_{-0.2}^{+0.2}$ & $8.8_{-0.4}^{+0.4}$ & $9.4_{-0.3}^{+0.3}$& $7.2_{-0.5}^{+0.5}$ & $6.7_{-0.2}^{+0.2}$& $5.8_{-0.2}^{+0.2}$ \\    
 	 \multicolumn{8}{c}{---Neutral iron lines---}\\
$F_{640}$&$1.03_{-0.08}^{+0.08}$ &$1.06_{-0.07}^{+0.05}$ &$1.52_{-0.16}^{+0.15}$ & $1.27_{-0.12}^{+0.11}$&$1.37_{-0.26}^{+0.26}$ & $1.71_{-0.08}^{+0.08}$ & $1.63_{-0.17}^{+0.18}$ \\
$F_{706}$ & $0.13_{-0.07}^{+0.07}$ &$0.15_{-0.15}^{+0.05}$&$0.16_{-0.14}^{+0.16}$ & $0.17_{-0.09}^{+0.09}$& $0.10_{-0.10}^{+0.32}$ & $0.17_{-0.04}^{+0.09}$ &$0.22_{-0.13}^{+0.15}$ \\
      \multicolumn{8}{c}{---Highly ionized iron lines---}\\
$F_{667}$ & $0.91_{-0.08}^{+0.04}$ &$0.89_{-0.06}^{+0.06}$ &$1.22_{-0.16}^{+0.15}$ & $0.81_{-0.11}^{+0.11}$& $1.14_{-0.27}^{+0.27}$ & $1.00_{-0.07}^{+0.08}$& $1.20_{-0.18}^{+0.18}$ \\
      \hline
$\delta_{\rm Gain}$\footnotemark[$a$] & 0(fix) & $21_{-3}^{+3}$ & 0(fix) & $35_{-11}^{+4}$ & 0(fix) & $-20_{-3}^{+4}$& $71_{-13}^{+21}$ \\
$\chi^2$/dof &88.9/93 & 185.6/169 & 84.2/101 & 260.4/164 &37.5/50 & 196.0/169 & 81.8/76 \\
      \hline
        \multicolumn{8}{@{}l@{}}{\hbox to 0pt{\parbox{160mm}{\footnotesize
Note---Terminologies and units are same as table 2.\\
The intensity of Fe \emissiontype{XXVI} K$\alpha$ is fixed at 30\% of that of Fe \emissiontype{XXV} K$\alpha$ determined by the Suzaku fit.\\
The other parameters are fixed to the best-fit Suzaku values.		
\par\noindent
\footnotemark[$a$] Fine-tuning offset energy in units of eV.
      	\par\noindent
         }\hss}}
    \end{tabular}
  \end{center}
\end{table*}

\subsection{Systematic Errors of the Line Flux}

The estimation of the line flux depends on 
the underlying continuum shape and the background subtraction.
We thus examined the possible systematic error cause by 
the assumed continuum and background subtraction.  

\subsubsection{Dependence of the Line Flux on the Continuum Shape}

We fixed either $\Gamma$ (Case 2) or $N_{\rm H}$ (Case 3) 
or both (Case 4) to those of the best-fit Suzaku results 
and fitted the XMM-Newton, Chandra and ASCA spectra.
For comparison, we call fitting 
with free $\Gamma$ and $N_{\rm H}$ as Case 1.
The best-fit parameters of these fittings (Cases 2, 3, and 4) 
are shown in table 4.

\begin{table*}
  \caption{Dependence of the best-fit flux on the continuum parameters for the Sgr B2 region}
  \label{tab:specfit_b2area_asca}
  \begin{center}
    \begin{tabular}{llllll}
      \hline
&Parameter & Case1&  Case2& Case3& 	Case4 \\
	\multicolumn{6}{c}{---XMM-Newton---} \\
MOS(2004)&$F_{\rm pow}$ & $4.8_{-0.2}^{+0.2}$ & $4.9_{-0.2}^{+0.2}$ & $4.8_{-0.2}^{+0.2}$ & $4.8_{-0.2}^{+0.2}$\\ 
&$F_{640}$ & $1.03_{-0.08}^{+0.08}$ & $1.04_{-0.08}^{+0.08}$ & $1.07_{-0.07}^{+0.07}$ & $1.07_{-0.07}^{+0.07}$ \\
&$F_{706}$ & $0.13_{-0.07}^{+0.07}$ & $0.10_{-0.07}^{+0.07}$ & $0.12_{-0.07}^{+0.07}$ & $0.12_{-0.07}^{+0.07}$\\ 
&$F_{667}$ & $0.91_{-0.08}^{+0.04}$ & $0.92_{-0.04}^{+0.08}$ & $0.95_{-0.07}^{+0.07}$ & $0.95_{-0.07}^{+0.07}$ \\
PN(2004)&$F_{\rm pow}$ & $7.0_{-0.2}^{+0.2}$ & $6.8_{-0.2}^{+0.2}$ & $7.1_{-0.2}^{+0.2}$ & $6.6_{-0.2}^{+0.2}$ \\
&$F_{640}$ & $1.06_{-0.07}^{+0.05}$ & $1.02_{-0.05}^{+0.08}$ & $1.01_{-0.05}^{+0.07}$ & $1.07_{-0.06}^{+0.06}$ \\
&$F_{706}$ & $0.15_{-0.15}^{+0.05}$ & $0.06_{-0.06}^{+0.05}$ & $0.10_{-0.10}^{+0.05}$ & $0.08_{-0.05}^{+0.04}$ \\
&$F_{667}$ & $0.89_{-0.06}^{+0.06}$ & $0.88_{-0.09}^{+0.05}$ & $0.85_{-0.06}^{+0.05}$ & $0.92_{-0.05}^{+0.05}$ \\
MOS(2001)&$F_{\rm pow}$ & $8.8_{-0.4}^{+0.4}$ & $8.2_{-0.3}^{+0.3}$ & $8.9_{-0.4}^{+0.4}$ & $7.7_{-0.3}^{+0.3}$\\ 
&$F_{640}$ & $1.52_{-0.16}^{+0.15}$ & $1.53_{-0.26}^{+0.12}$ & $1.43_{-0.16}^{+0.15}$ & $1.55_{-0.16}^{+0.14}$\\
&$F_{706}$ & $0.16_{-0.14}^{+0.16}$ & $0.31_{-0.22}^{+0.12}$ & $0.16_{-0.16}^{+0.14}$ & $0.32_{-0.14}^{+0.14}$ \\
&$F_{667}$ & $1.22_{-0.16}^{+0.15}$ & $1.16_{-0.19}^{+0.17}$ & $1.10_{-0.15}^{+0.15}$ & $1.28_{-0.14}^{+0.14}$ \\
PN(2001)&$F_{\rm pow}$ &$9.4_{-0.3}^{+0.3}$ & $9.2_{-0.3}^{+0.3}$ & $9.6_{-0.3}^{+0.3}$ & $8.6_{-0.3}^{+0.3}$ \\
&$F_{640}$ & $1.27_{-0.12}^{+0.11}$ & $1.19_{-0.11}^{+0.11}$ & $1.21_{-0.11}^{+0.11}$ & $1.29_{-0.10}^{+0.11}$ \\
&$F_{706}$& $0.17_{-0.09}^{+0.09}$ & $0.19_{-0.08}^{+0.09}$ & $0.14_{-0.08}^{+0.10}$ & $0.28_{-0.08}^{+0.09}$ \\
&$F_{667}$& $0.81_{-0.11}^{+0.11}$ & $0.72_{-0.10}^{+0.10}$ & $0.73_{-0.10}^{+0.10}$ & $0.88_{-0.10}^{+0.09}$ \\	
	\multicolumn{6}{c}{---Chandra---} \\
ACIS(2001)&$F_{\rm pow}$ & $7.2_{-0.5}^{+0.5}$ & $6.6_{-0.5}^{+0.5}$ & $7.1_{-0.5}^{+0.5}$ & $6.0_{-0.4}^{+0.4}$ \\
&$F_{640}$ & $1.37_{-0.26}^{+0.26}$ & $1.36_{-0.25}^{+0.27}$ & $1.33_{-0.24}^{+0.27}$ & $1.48_{-0.24}^{+0.24}$\\
&$F_{706}$ & $0.10_{-0.10}^{+0.32}$ & $0.12_{-0.12}^{+0.28}$ & $0.12_{-0.12}^{+0.32}$ & $0.20_{-0.20}^{+0.28}$ \\
&$F_{667}$ & $1.14_{-0.27}^{+0.27}$ & $1.16_{-0.27}^{+0.27}$ & $1.13_{-0.41}^{+0.14}$ & $1.30_{-0.24}^{+0.24}$ \\
ACIS(2000)&$F_{\rm pow}$ & $6.7_{-0.2}^{+0.2}$ & $6.0_{-0.2}^{+0.2}$ & $6.5_{-0.2}^{+0.2}$ & $5.7_{-0.2}^{+0.2}$ \\
&$F_{640}$& $1.71_{-0.08}^{+0.08}$ & $1.72_{-0.10}^{+0.07}$ & $1.67_{-0.08}^{+0.08}$ & $1.76_{-0.08}^{+0.08}$ \\
&$F_{706}$& $0.17_{-0.04}^{+0.09}$ & $0.26_{-0.08}^{+0.12}$ & $0.21_{-0.09}^{+0.08}$ & $0.31_{-0.09}^{+0.09}$ \\
&$F_{667}$ & $1.00_{-0.07}^{+0.08}$ & $1.02_{-0.08}^{+0.09}$ & $0.97_{-0.07}^{+0.08}$ & $1.08_{-0.08}^{+0.07}$ \\
	\multicolumn{6}{c}{---ASCA---} \\
SIS(1994)&$F_{\rm pow}$ & $5.8_{-0.2}^{+0.2}$ & $5.5_{-0.2}^{+0.2}$ & $5.8_{-0.2}^{+0.2}$ & $5.4_{-0.2}^{+0.2}$ \\
&$F_{640}$ & $1.63_{-0.17}^{+0.18}$ & $1.65_{-0.18}^{+0.16}$ & $1.62_{-0.19}^{+0.14}$ & $1.65_{-0.10}^{+0.17}$ \\
&$F_{706}$ & $0.22_{-0.13}^{+0.15}$ & $0.30_{-0.15}^{+0.17}$ & $0.24_{-0.15}^{+0.13}$ & $0.30_{-0.14}^{+0.17}$ \\
&$F_{667}$ & $1.20_{-0.18}^{+0.18}$ & $1.14_{-0.18}^{+0.16}$ & $1.07_{-0.13}^{+0.18}$ & $1.15_{-0.16}^{+0.15}$ \\
	\hline
       \multicolumn{6}{@{}l@{}}{\hbox to 0pt{\parbox{120mm}{\footnotesize
      	Note---Same as table 3. \\
Case 1: $N_{\rm H}$ and $\Gamma$ are free parameters (same as table 2 and 3) \\
Case 2: $N_{\rm H}$ is a free parameter but $\Gamma$ is fixted to the best-fit Suzaku value. \\
Case 3: $\Gamma$ is a free parameter but $N_{\rm H}$ is fixted to the best-fit Suzaku value. \\
Case 4: $N_{\rm H}$ and $\Gamma$ are fixted to the best-fit Suzaku value. \\
    }\hss}}
    \end{tabular}
  \end{center}
\end{table*}

From table 4, we find that the best-fit fluxes of 
the 6.40~keV and 6.67~keV lines for the four fitting cases 
are consistent within $\pm$5\% with each other, 
which is smaller than the statistical errors.
We therefore ignore these systematic errors, 
and refer to the Case 1 parameters in the subsequent discussion.

\subsubsection{Dependence of the Line Flux on the Background Subtraction}

As we have already noted, the NXB of Chandra 
in the 2001 observation was unusually high.
This unstable (time-dependent) NXB is also often the problem 
in XMM-Newton analysis \citep{Carter2007}.
The unstable NXB and associated lines may affect 
the estimated line fluxes in the source spectra.
To check these systematic errors, we derived the count rates 
of MOS and PN for the source and background above the 10~keV band, 
where NXB is dominant.
The count-rate ratios between the source and background 
were used as correction factors $\alpha$ for the background.
The factors $\alpha$ for the XMM-Newton data in 2004 
are relatively small, 1.0 and 1.4 for MOS and PN, respectively, 
while those for the data in 2001 
are 2.1 and 2.6 for MOS and PN, respectively.

We multiplied the background spectra by the factors $\alpha$, 
subtracted them from the source spectra, 
and then fitted them with the model expressed by equation (1) 
for the four cases (Case 1 to 4). 
The results are shown in table 5.
In this process, the CXB was also multiplied by the factor $\alpha$.
Therefore, we over(under)-subtracted the CXB by the factor $\alpha$.
The surface brightness of the CXB with a typical Galactic absorption 
of 6$\times 10^{22}$ cm$^{-2}$ is 
$\sim 10^{-15}$ ergs cm$^{-2}$ s$^{-1}$ arcmin$^{-2}$ 
in the Sgr B2 region, 
which is only a few percent of the continuum flux in the region.
Thus, possible over(under)-subtraction of the CXB 
can be neglected in the following discussion.

From tables 4 and 5, we find that the fluxes of 
the 6.40~keV and 6.67~keV lines with corrected background subtraction 
(table 5) are almost the same as those with normal background 
subtraction (table 4) within the statistical errors, 
although the continuum flux is substantially different.
Thus, in the following discussion based on 
the fluxes of the 6.40~keV and 6.67~keV lines, 
we can ignore possible uncertainty 
in the time-dependent background levels.  

\begin{table*}
  \caption{Dependence of the XMM-Newton flux on the background-subtraction for the Sgr B2 region}
  \label{tab:specfit_b2area_xmm_multinxb}
  \begin{center}
    \begin{tabular}{lllll}
      \hline\hline
Parameter & Case1& 	Case2& 	Case3& 	Case4 \\
	\multicolumn{5}{c}{XMM-Newton/PN(2004): Correction factor = 1.4} \\
$F_{\rm pow}$& $5.5_{-0.2}^{+0.2}$ & $5.6_{-0.2}^{+0.2}$ & $5.5_{-0.2}^{+0.2}$ & $5.5_{-0.2}^{+0.2}$ \\
$F_{640}$ &$1.05_{-0.06}^{+0.06}$ & $1.07_{-0.06}^{+0.06}$ & $1.08_{-0.06}^{+0.06}$ & $1.08_{-0.06}^{+0.06}$ \\
$F_{706}$ & $0.06_{-0.05}^{+0.05}$ & $0.06_{-0.05}^{+0.06}$ & $0.07_{-0.05}^{+0.05}$ & $0.07_{-0.06}^{+0.04}$ \\
$F_{667}$ & $0.90_{-0.06}^{+0.05}$ & $0.91_{-0.06}^{+0.06}$ & $0.92_{-0.06}^{+0.06}$ & $0.92_{-0.06}^{+0.06}$ \\
$\chi^2$/dof & 238.4/169 & 240.9/170 & 240.7/170 & 241.1/171 \\
      	\multicolumn{5}{c}{XMM-Newton/MOS(2001): Correction factor = 2.1} \\
$F_{\rm pow}$ & $3.9_{-0.3}^{+0.3}$ & $4.4_{-0.3}^{+0.3}$ & $4.1_{-0.3}^{+0.3}$ & $4.6_{-0.3}^{+0.3}$ \\
$F_{640}$ & $1.49_{-0.16}^{+0.16}$ & $1.51_{-0.16}^{+0.16}$ & $1.52_{-0.15}^{+0.16}$ & $1.46_{-0.15}^{+0.15}$\\
$F_{706}$ & $0.21_{-0.15}^{+0.15}$ & $0.14_{-0.14}^{+0.14}$ & $0.17_{-0.14}^{+0.15}$ & $0.10_{-0.10}^{+0.14}$\\
$F_{667}$ & $1.23_{-0.16}^{+0.16}$ & $1.23_{-0.16}^{+0.15}$ & $1.25_{-0.15}^{+0.15}$ & $1.17_{-0.15}^{+0.15}$ \\
$\chi^2$/dof & 84.2/101 & 92.6/102 & 88.0/102 & 94.7/103 \\
	\multicolumn{5}{c}{XMM-Newton/PN(2001): Correction factor = 2.6} \\
$F_{\rm pow}$ & $5.0_{-0.3}^{+0.3}$ &$5.1_{-0.3}^{+0.3}$ & $4.9_{-0.3}^{+0.3}$ & $5.0_{-0.3}^{+0.3}$ \\
$F_{640}$& $1.28_{-0.11}^{+0.12}$ & $1.32_{-0.12}^{+0.11}$ & $1.35_{-0.06}^{+0.03}$ & $1.34_{-0.11}^{+0.11}$\\
$F_{706}$& $0.24_{-0.10}^{+0.09}$ & $0.21_{-0.09}^{+0.10}$ & $0.24_{-0.03}^{+0.05}$ & $0.23_{-0.09}^{+0.10}$ \\
$F_{667}$& $0.84_{-0.11}^{+0.11}$ & $0.87_{-0.05}^{+0.11}$ & $0.92_{-0.03}^{+0.04}$ & $0.90_{-0.10}^{+0.09}$ \\
$\chi^2$/dof & 260.4/164 & 267.5/165 & 267.1/165 & 267.9/166 \\
      	\hline
        \multicolumn{5}{@{}l@{}}{\hbox to 0pt{\parbox{80mm}{\footnotesize
      	Note---Same as table 4.
      }\hss}}
    \end{tabular}
  \end{center}
\end{table*}

\subsection{The Constant 6.67~keV Line Flux}

The most serious problem when comparing the line fluxes 
between different satellites or different instruments 
is the reliability of the cross-calibration of their relative efficiencies.
The 6.67~keV line emission is due to the
largely extended GCDX and a new SNR G\,0.61+0.01,
which contributes about 30\% of the total 6.67~keV line emission 
in the Sgr B2 region.
Therefore, the 6.67~keV line flux is time invariant and
hence is a good cross-calibration line 
for the relative efficiencies 
of the Suzaku, XMM-Newton, Chandra, and ASCA observations.  
The constant flux hypothesis for the best-fit 6.67~keV line 
given in tables 2 and 3 is rejected with $\chi^2$/dof=56.3/8,
which indicates that there are significant systematic errors 
in the detection efficiencies of 
the ASCA, XMM-Newton, Chandra and Suzaku observations.
 
\begin{figure}
  \begin{center}
    \FigureFile(80mm,50mm){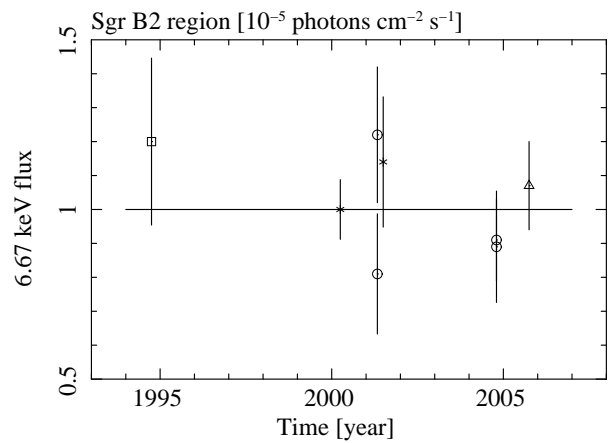}
 \end{center}
\caption{The time trend of 6.67~keV line fluxes in the Sgr B2 region.
  In this plot, each error bar indicates one sigma
  including the statistical and systematic uncertainties.
  The Suzaku data are indicated by triangles, 
  XMM-Newton MOS data by circles, 
  Chandra data by asterisks, and ASCA data by squares. 
  The average flux is shown by the solid horizontal line.}
\label{fig:trend_b2area_6.67_sys}
\end{figure}

The nominal systematic errors for these satellites are: 
13\% for the ASCA SIS\footnote{See http://heasarc.gsfc.nasa.gov/listserv/ascanews/msg00143.html},
5\% for Chandra \citep{Schwartz2000}, 
10\% for EPIC 
(also the MOS flux is 10--15\% higher than the PN \citep{Carter2007}) 
and 10\% for Suzaku \citep{Serlemitosos2007}.

The 6.67~keV line flux history obtained by adding these systematic errors 
to the statistical errors are plotted 
in figure \ref{fig:trend_b2area_6.67_sys}.
The $\chi^2$/dof value with a constant flux model is then 4.9/8, 
which is acceptable at a 70\% confidence level.
Therefore, as expected, the 6.67~keV line flux is constant, 
if systematic errors are taken into account.

\begin{figure}
  \begin{center}
    \FigureFile(80mm,50mm){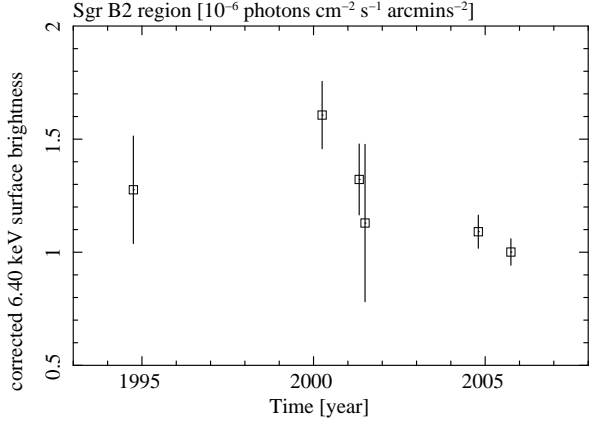}
  \end{center}
\caption{The light curve of 
  the corrected 6.40~keV line surface brightness in the Sgr B2 region.
  In this plot, each error bar indicates a 90\% confidence limit.}
\label{fig:trend_b2area_6.40_cor}
\end{figure}

\subsection{Time-Variable 6.40 keV Line}

Unlike the 6.67~keV line, the 6.40~keV line emission is 
much more clumpy, hence it may not be time constant. 
To examine whether the 6.40~keV line from the Sgr B2 region 
is time variable or not,
we performed a $\chi^2$ test using the same procedure as 
those of the 6.67~keV line.
For the case with no systematic errors, 
the constant flux hypothesis is rejected with $\chi^2$/dof = 112.8/8.

We next added the systematic errors in the same way as the 6.67~keV line case,
and the constant flux hypothesis was still rejected 
with $\chi^2$/dof = 17.9/8.
Therefore, the 6.40~keV line flux is time variable.

Having confirmed that the 6.67~keV line is constant, 
we used it as a cross-calibrator.
We introduced normalization factors for the detection efficiencies 
($c\pm \delta c$) for the eight observations 
shown in table \ref{tab:67normalizefactors}, 
where $\delta c$ is a statistical error. 
These parameters are the ratios of the best-fit 6.67~keV line flux 
of each satellite to the averaged flux. 

\begin{table*}
  \caption{The normalization factors of the detection efficiency for the eight observations}
  \label{tab:67normalizefactors}
  \begin{center}
    \begin{tabular}{cccccccc}
      \hline\hline
      Suzaku & \multicolumn{2}{c}{XMM-Newton} & \multicolumn{2}{c}{XMM-Newton} & Chandra & Chandra & ASCA \\
      XIS(2005) & MOS(2004) & PN(2004) & MOS(2001) & PN(2001) & ACIS(2001) & ACIS(2000) & SIS(1994) \\
      \hline
      $1.05\pm 0.05$ & $0.91\pm 0.08$ & $0.89\pm 0.06$ & $1.25\pm 0.14$ & $0.83\pm 0.11$ & $1.12\pm 0.29$ & $0.99\pm 0.07$ & $1.19\pm 0.18$ \\
      \hline
\multicolumn{8}{@{}l@{}}{\hbox to 0pt{\parbox{160mm}{\footnotesize
Note---Parentheses indicate in the 90\% confidence limit.
\par\noindent
Normalization factors and their errors of the detection efficiency derived from the ratio of the best-fit 6.67~keV line flux in the Sgr B2 region obtained with each observation to the averaged flux.
}\hss}}
  \end{tabular}
  \end{center}
\end{table*}

The best-fit 6.40~keV line fluxes were divided by these factors, 
and are referred to as the corrected 6.40~keV line flux.
Since the MOS and PN fluxes should be consistent, 
we averaged these fluxes and errors. 
Figure \ref{fig:trend_b2area_6.40_cor} shows 
the corrected 6.40~keV line fluxes in each observation period.  
The 6.40~keV line flux varies significantly 
with $\chi^2$/dof = 28.9/6.
The flux of the 2005 observation is about 60\% that of 2000. 

\subsection{Sub-structures in the Sgr B2 region}

In order to examine the morphology history of the 6.40~keV line,
we made a surface brightness map of iron lines for each observation.
Since the energy resolutions of Chandra or ASCA were not sufficiently good 
to separate the 6.40~keV line from the 6.67~keV line,
all the images are in the 6--7~keV band.
For the continuum flux subtraction from the 6--7~keV band flux, 
we assumed a power-law continuum of the photon index of 2.6, 
the best-fit Suzaku value, and estimated that the power-law continuum 
in the 6--7 keV band is 65\% of the 5--6~keV band flux.
After subtracting the continuum emission, 
we made the surface brightness maps 
using a pixel unit of 50-arcsec square, 
similar to the PSF of ASCA, 
which had the worst PSF among the four satellites.

\begin{figure*}
  \begin{center}
    \FigureFile(80mm,50mm){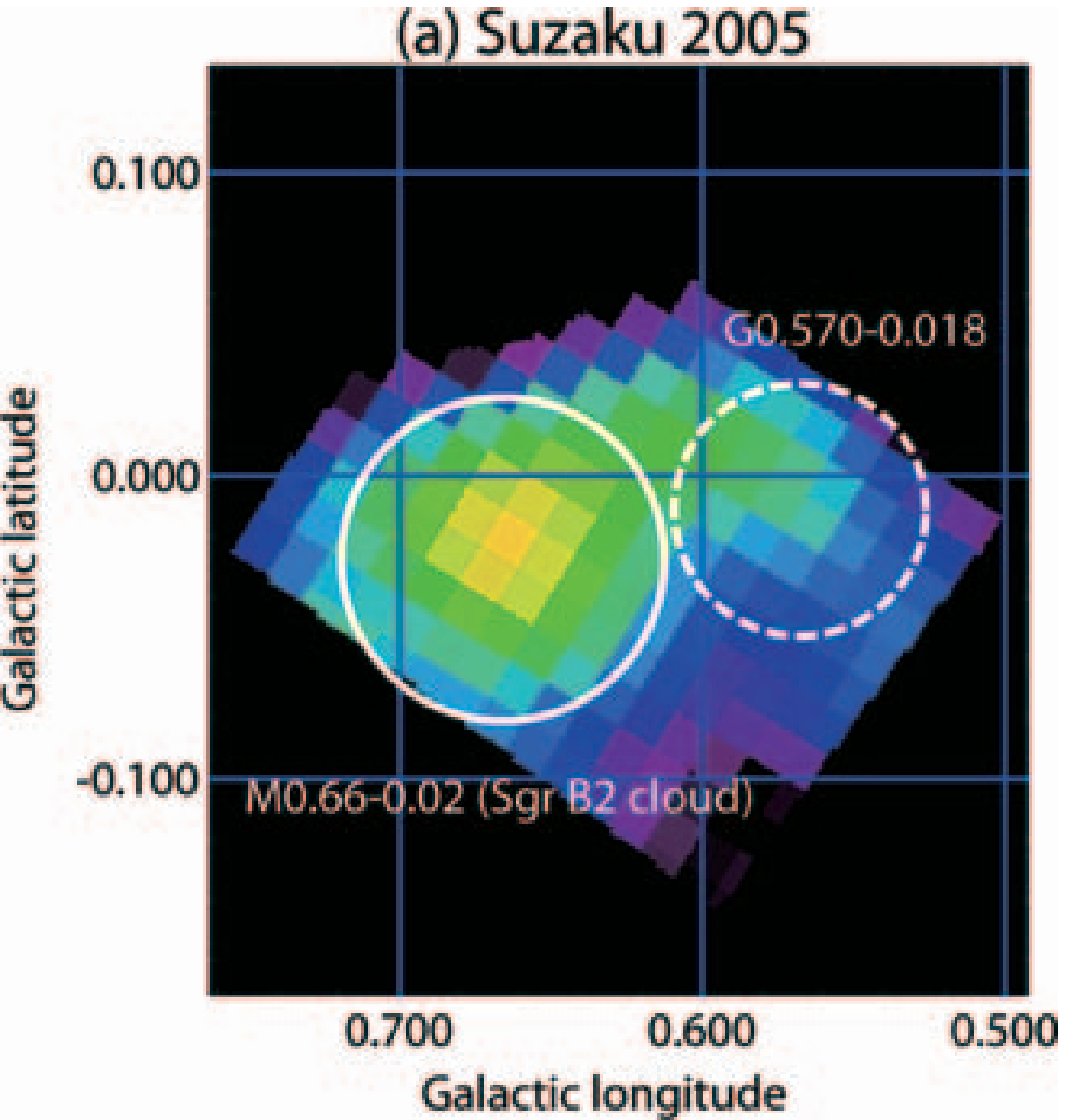}
    \FigureFile(80mm,50mm){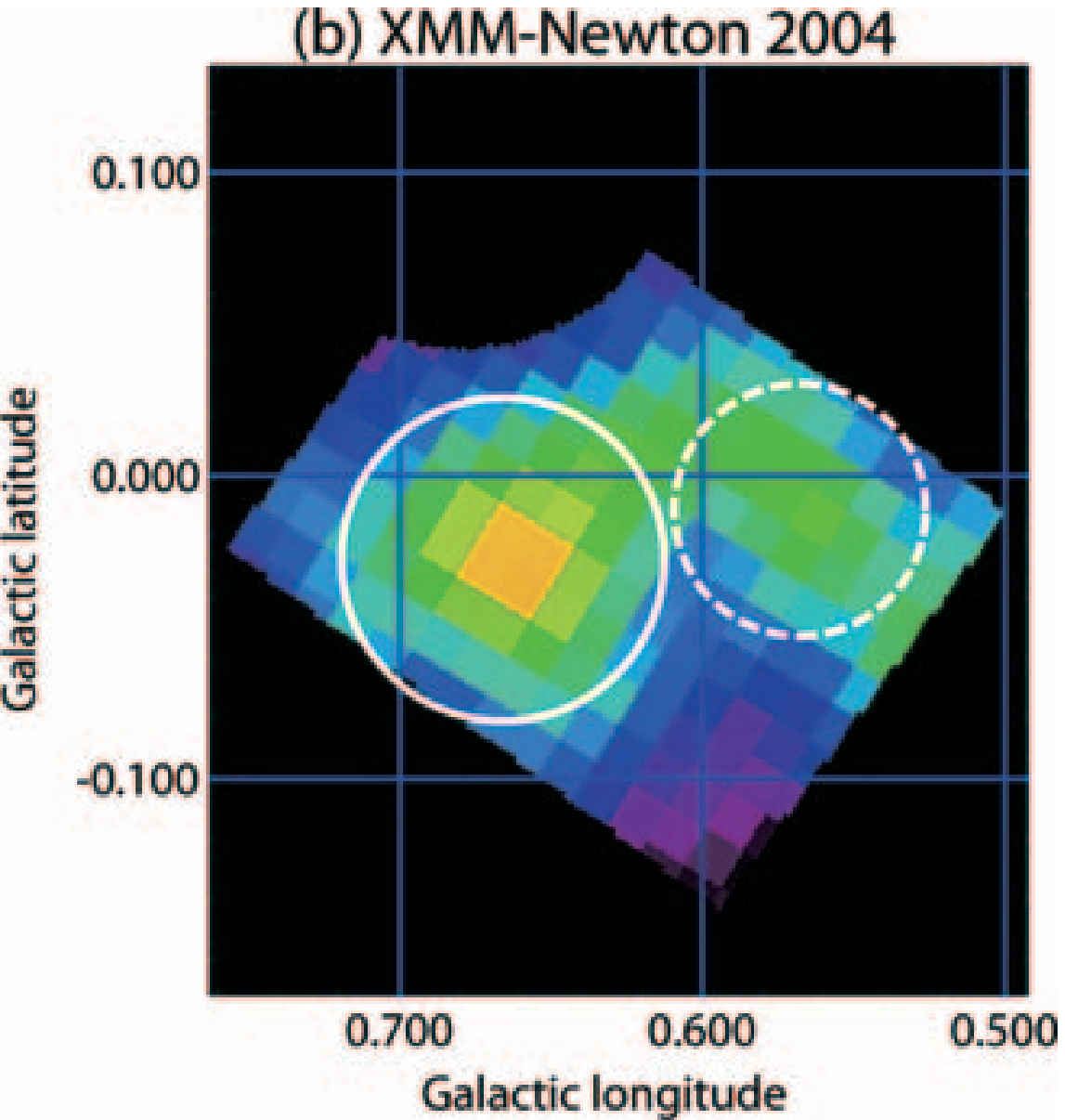}
    \FigureFile(80mm,50mm){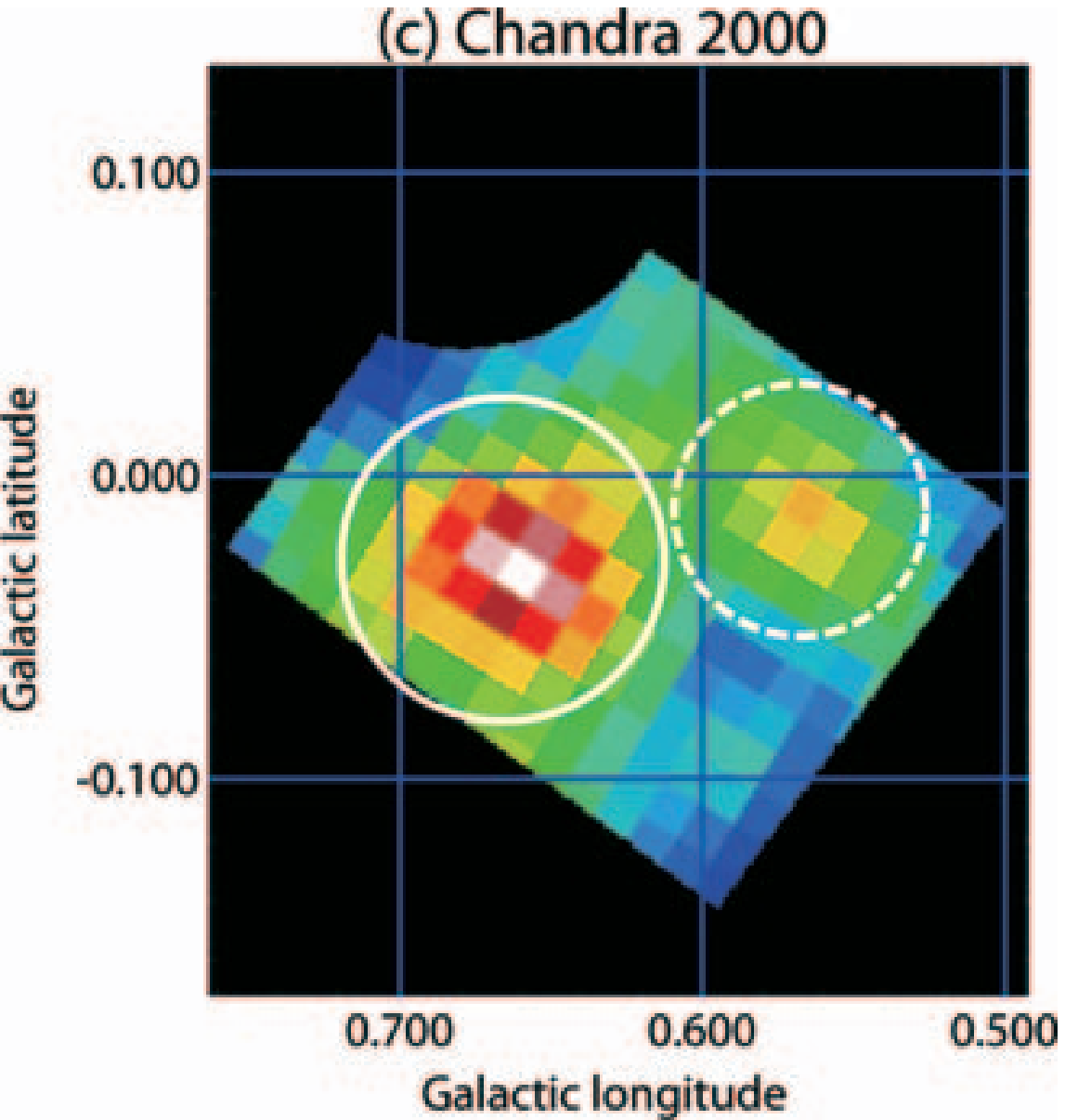}
    \FigureFile(80mm,50mm){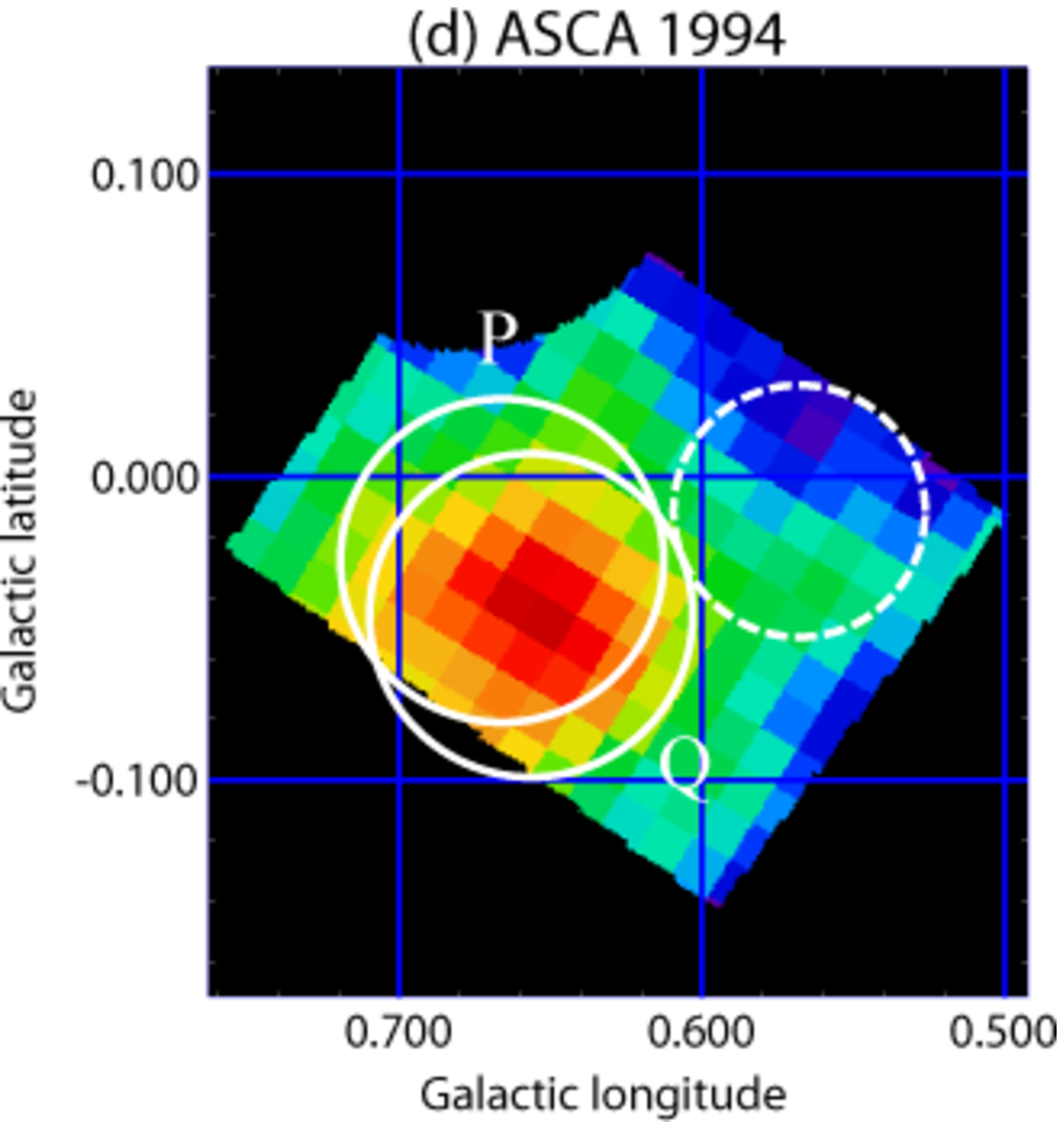}
  \end{center}
  \caption{Surface brightness maps of iron lines obtained with 
    (a) Suzaku XIS 2005, (b) XMM-Newton MOS and PN 2004, 
    (c) Chandra ACIS-I 2000, and (d) ASCA SIS 1994. 
    Pixel size is $\timeform{50''}\times\timeform{50''}$ in each case}
  \label{fig:img_ironsb}
\end{figure*}

Figures \ref{fig:img_ironsb}(a)--(d) show the surface brightness maps 
of the iron lines.
The Suzaku and XMM-Newton images are relatively dim 
in the whole area compared to Chandra and ASCA, 
which is consistent with the spectral fitting result.
In the Chandra image, a bright spot is found in the west, 
where a SNR candidate G\,0.570$-$0.018 is located \citep{Senda2002}.
We selected the brightest spot named M\,0.66$-$0.02 
(solid circle in figure \ref{fig:img_ironsb}) from its position, 
which was termed the ``Sgr B2 cloud'' in previous papers, 
and G\,0.570$-$0.018 (dashed circle in figure \ref{fig:img_ironsb}).
We then examined the 6.40~keV flux changes from these two spots.
Since the Chandra data in 2001 did not have enough statistics 
for the analysis,
we excluded this data hereafter. 

\subsubsection{M\,0.66$-$0.02}

We extracted the spectra of M\,0.66$-$0.02 (Sgr B2 cloud) 
from the solid circle with a radius of 3.2 arcmin 
in figure \ref{fig:img_ironsb}
and subtracted the blank-sky data in the same detector position.
In the ASCA image, the peak position of M\,0.66$-$0.02 
is different from those in the others.
It is not clear whether this shift is due to a relatively large error 
of the ASCA astrometry 
(no fine-tuning of the ASCA astrometry was performed, see Section 2.4) 
or due to a real shift in the variable 6.40~keV line.
Therefore, in the ASCA analysis, we selected two circles 
for M\,0.66$-$0.02, at the same as the position 
in the original coordinate of Suzaku, XMM-Newton, and Chandra (P) 
and at the center on the peak of the SIS image (Q).

\begin{figure*}
  \begin{center}
    \FigureFile(50mm,50mm){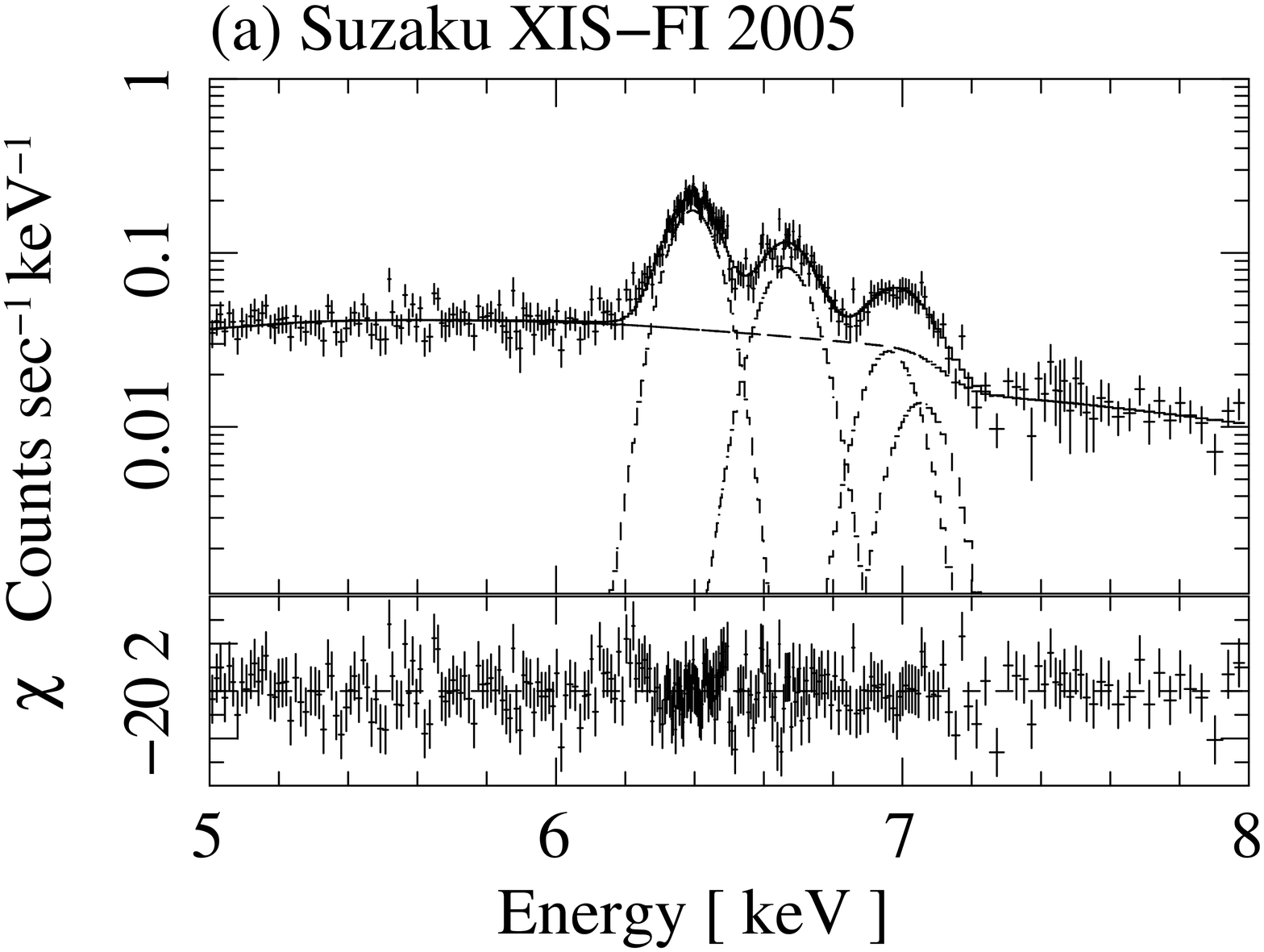}
    \FigureFile(50mm,50mm){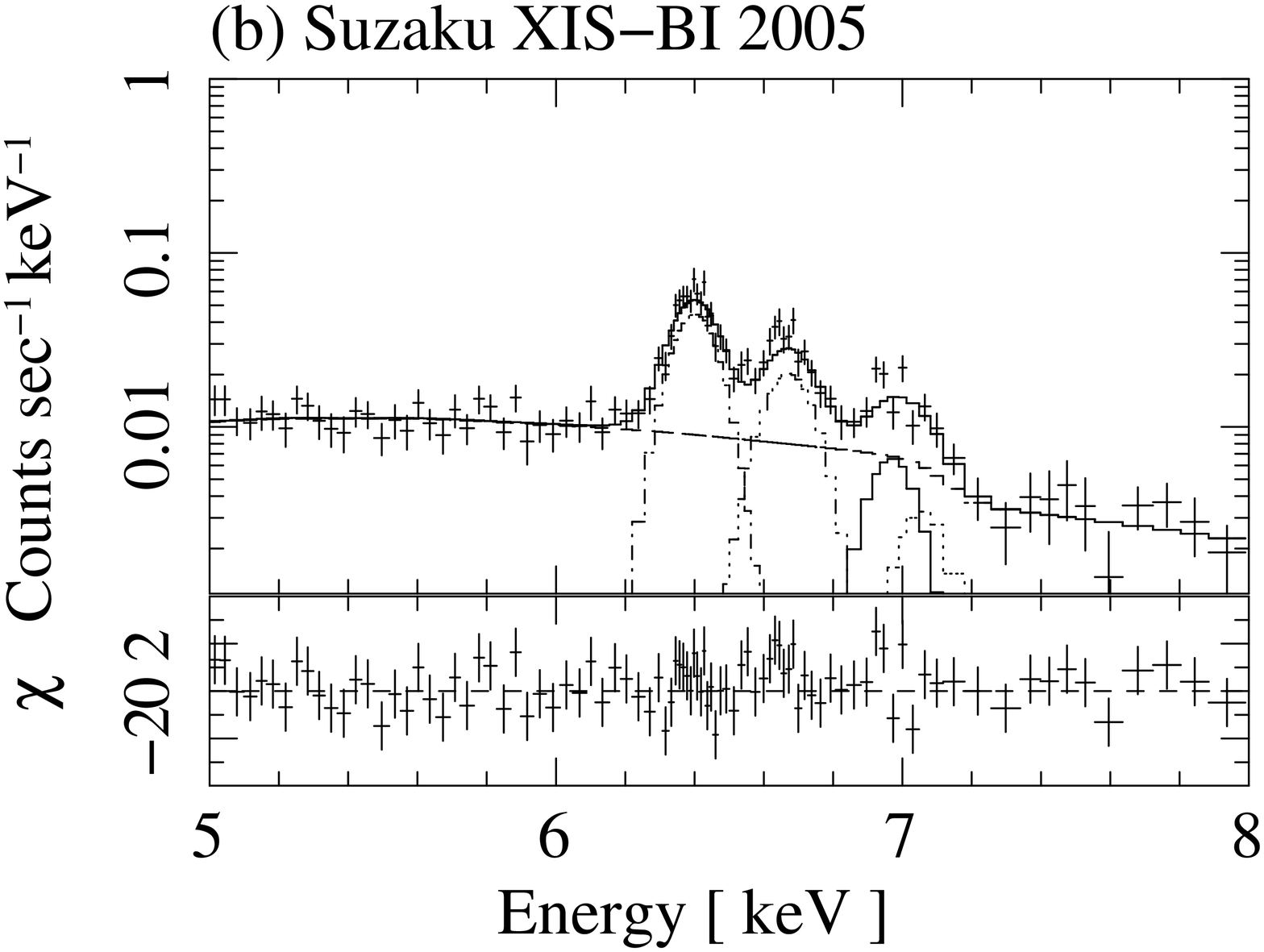}
    \FigureFile(50mm,50mm){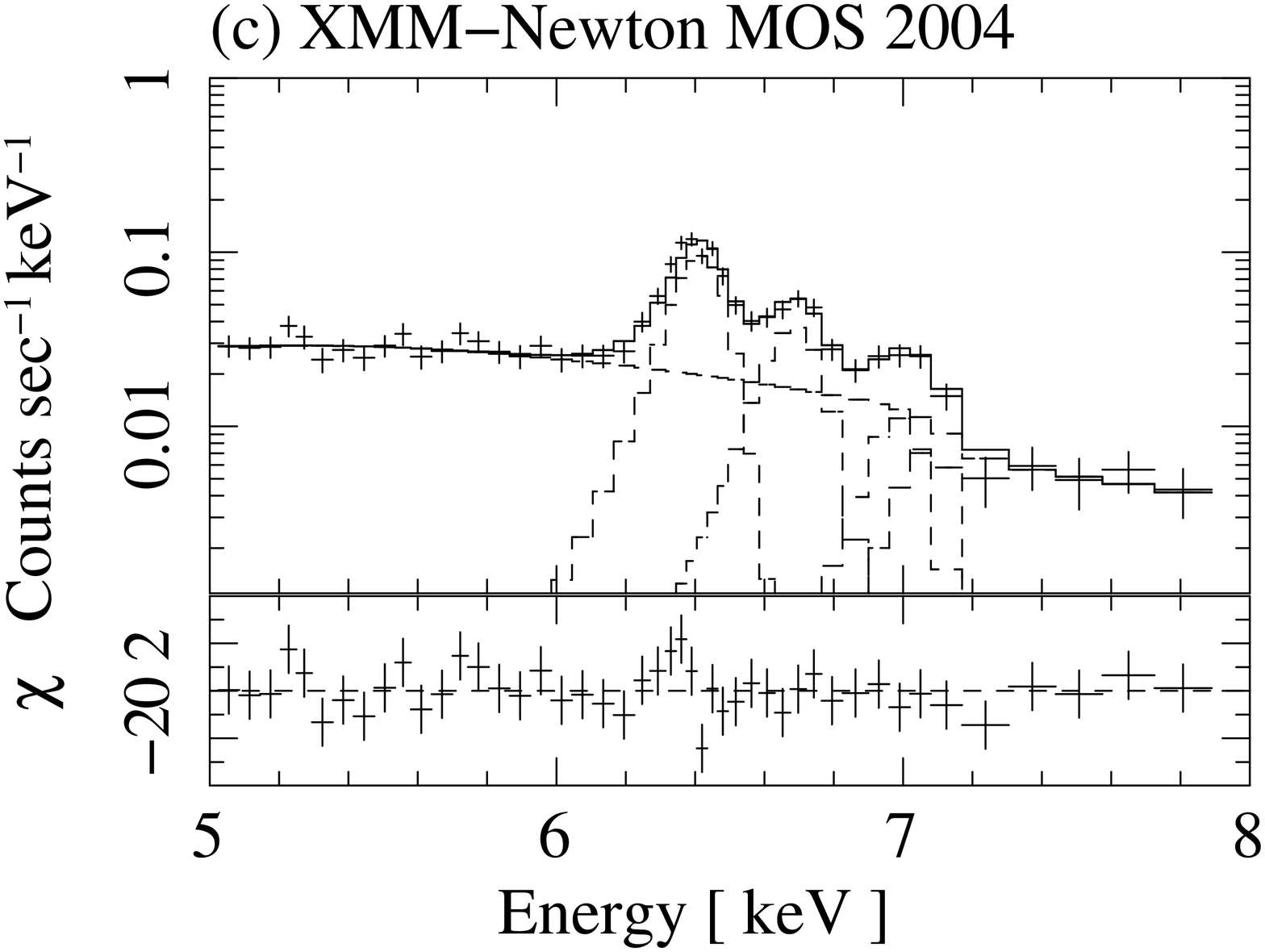}
    \FigureFile(50mm,50mm){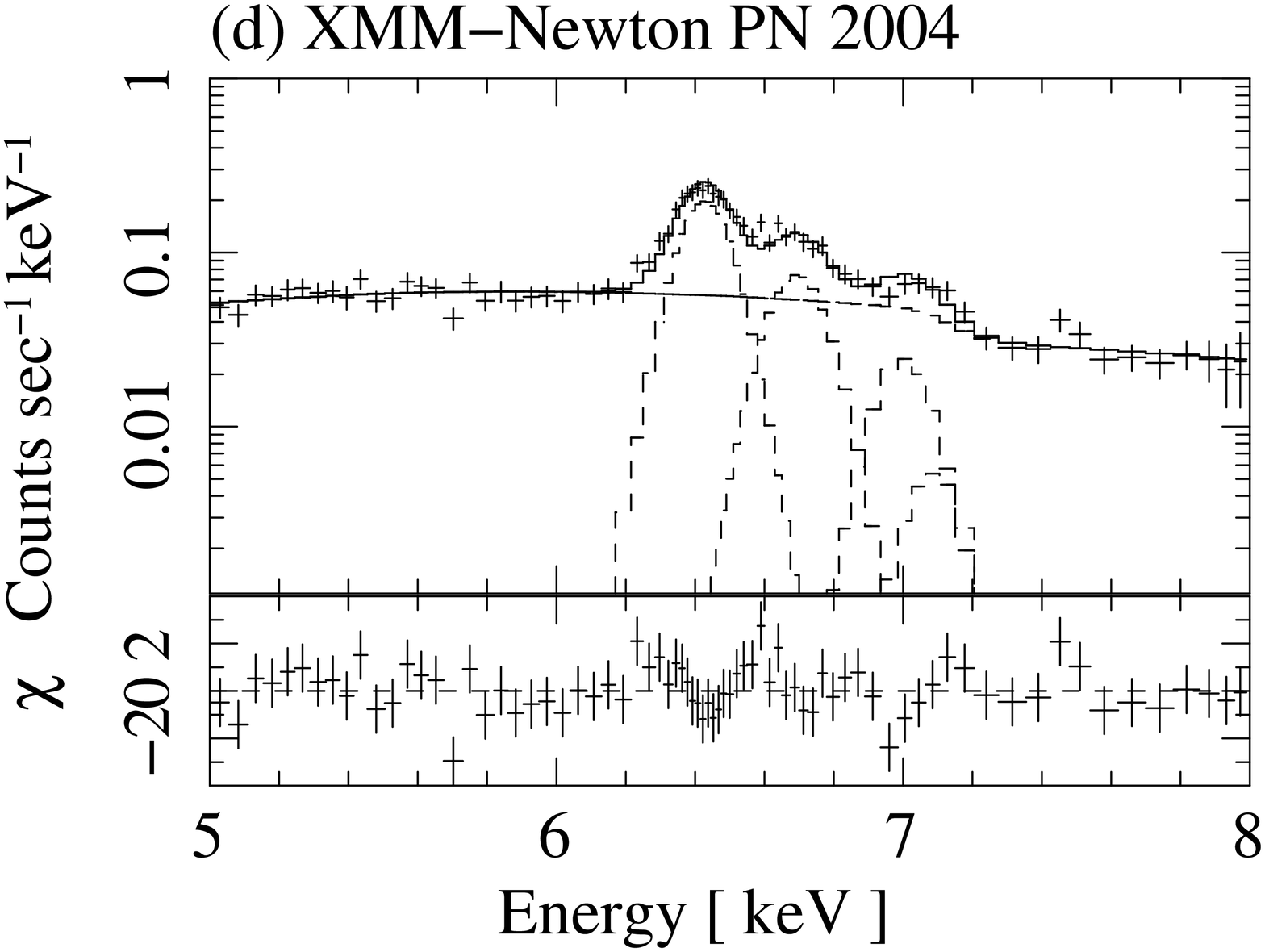}
    \FigureFile(50mm,50mm){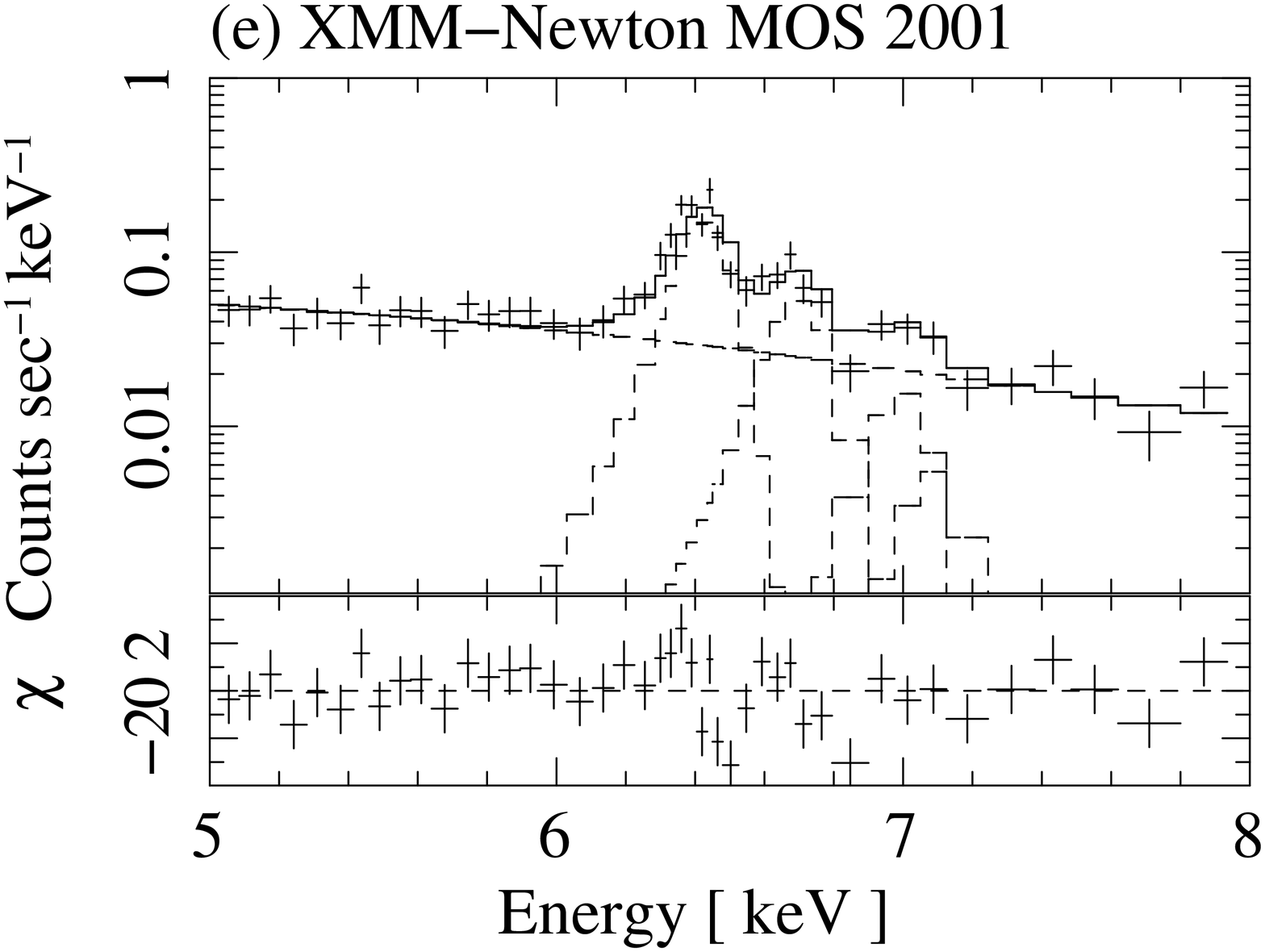}
    \FigureFile(50mm,50mm){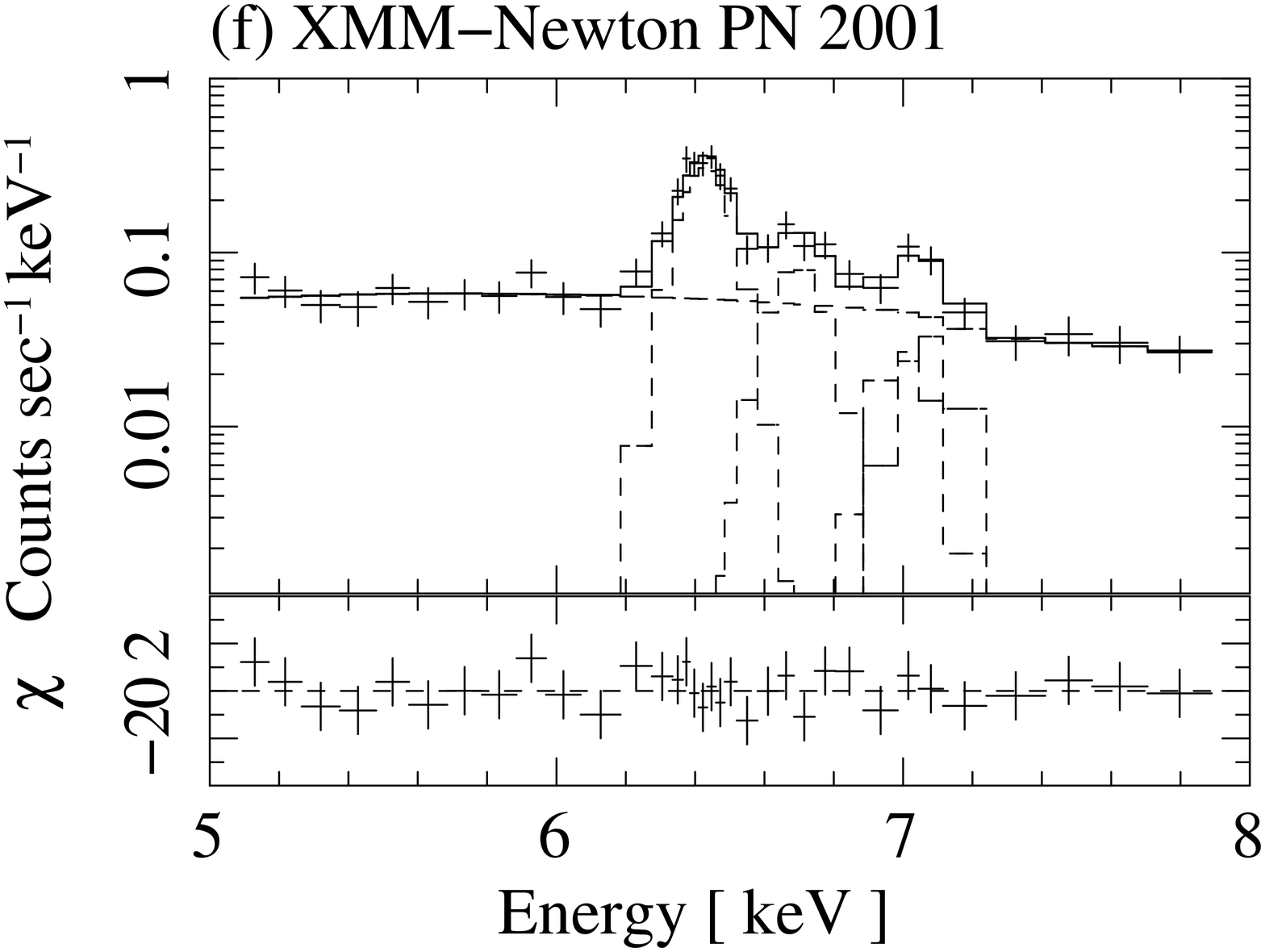}
    \FigureFile(50mm,50mm){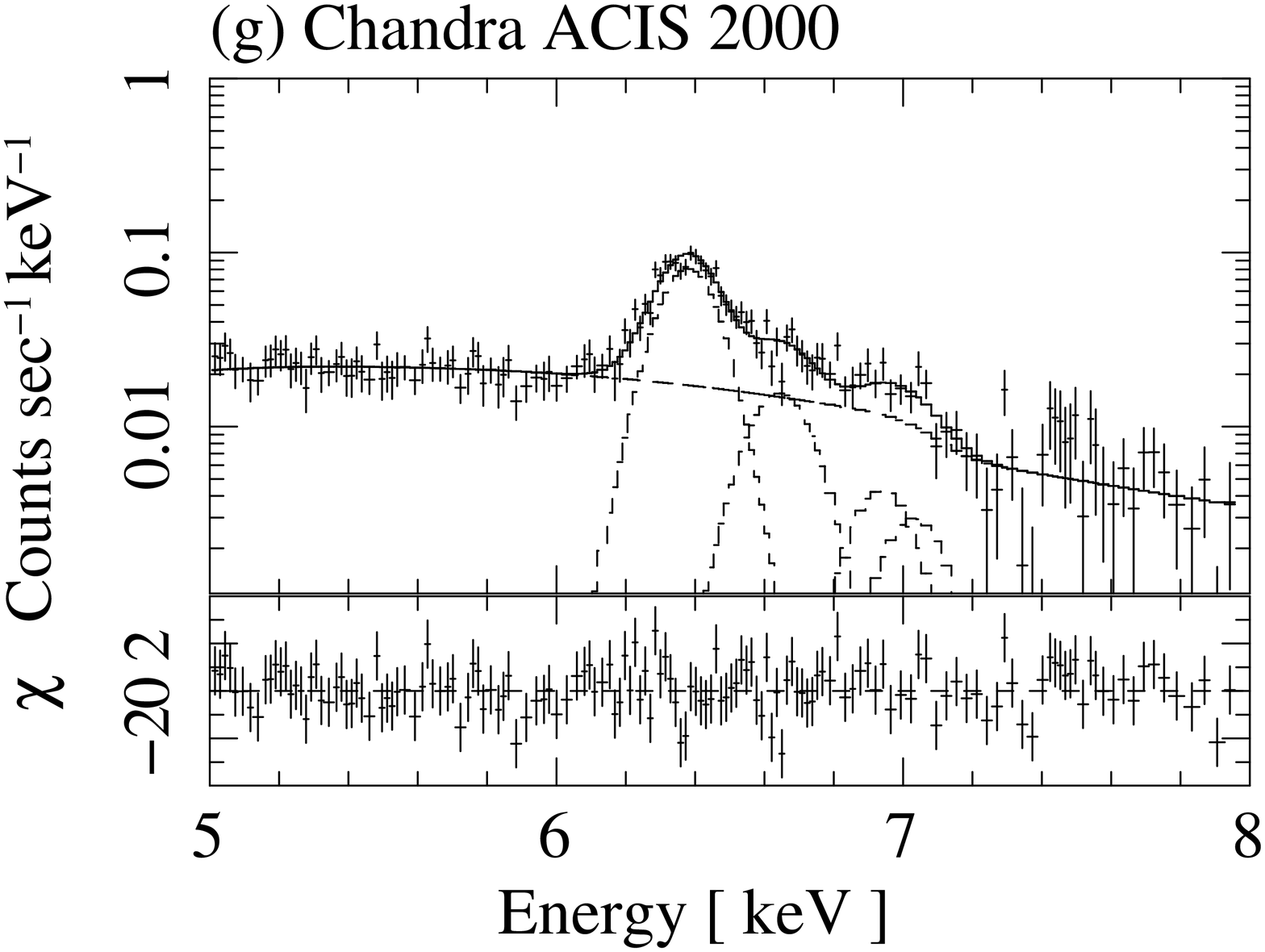}
    \FigureFile(50mm,50mm){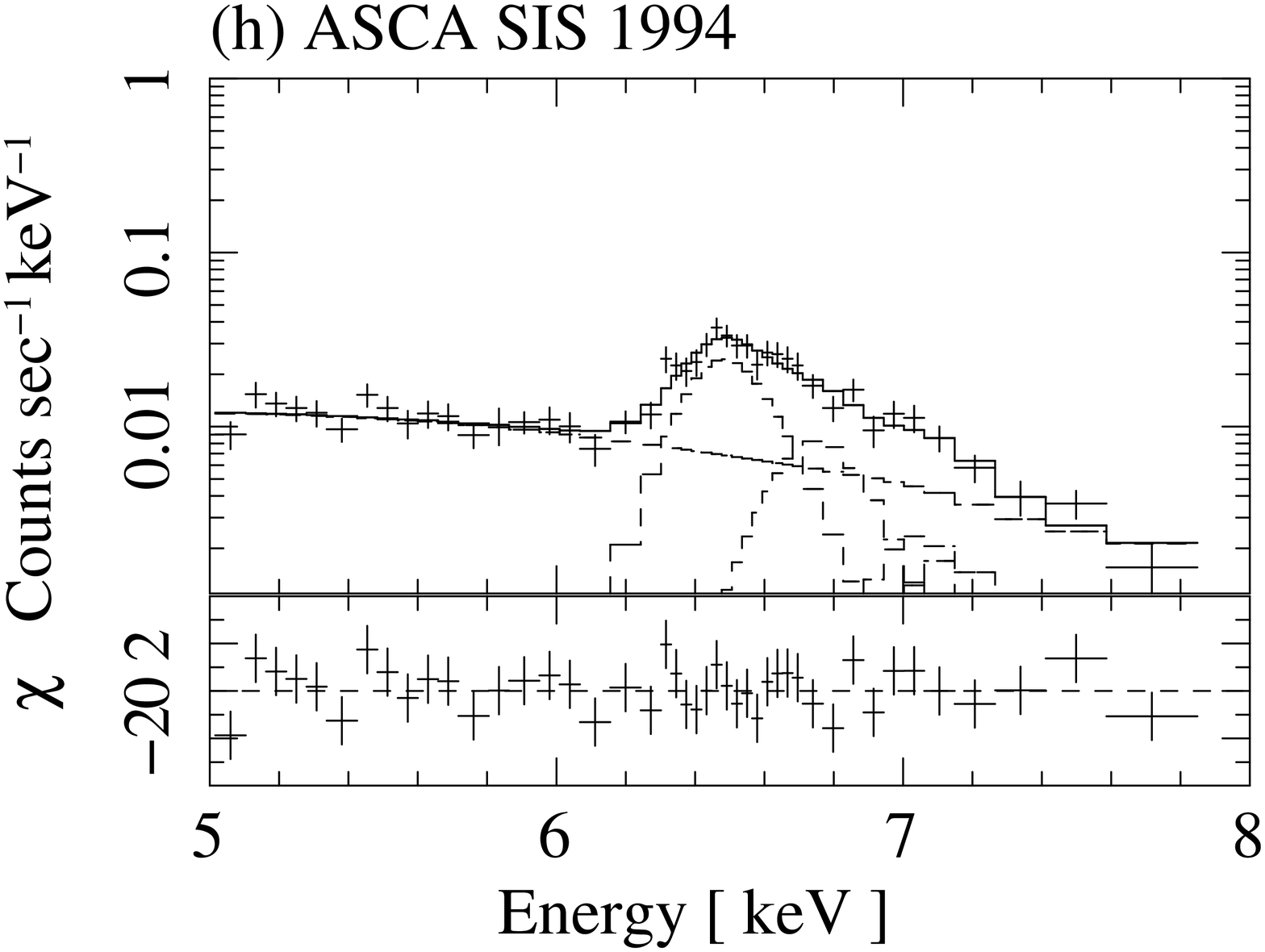}
  \end{center}
\caption{Spectra for M\,0.66$-$0.02 obtained with 
  Suzaku XIS FI sensors (a) and BI sensor (b) in 2005, 
  XMM-Newton MOS (c) and PN (d) in 2004, (e) and (f) in 2001, 
  Chandra ACIS-I (g) in 2000, and ASCA SIS (h) in 1994. 
  The solid lines indicate the best-fit models.}
\label{fig:spec_b2main2}
\end{figure*}

As described in Section 3.2, the dependences of the line fluxes 
on the continuum models and on the uncertainties of NXB 
(systematic errors) are not larger than the statistical errors 
of the Sgr B2 region.
Since the statistical error of the line flux from M\,0.66$-$0.02 
is larger than that of the Sgr B2 region,
these systematic errors can be neglected.
We nevertheless attempted the same analysis as the Sgr B2 region 
and found no difference within different continuum modeling 
and/or background subtractions.
We therefore estimated the line fluxes using the continuum models 
of $N_{\rm H}$ and $\Gamma$ free. 
The ASCA/SIS fluxes of the 6.40~keV line were $0.99_{-0.14}^{+0.13}$ 
for the spectrum region 'P',
and $0.93_{-0.15}^{+0.11}$ $\times 10^{-4}$ photons cm$^{-2}$ s$^{-1}$ 
for that of region 'Q'.
These fluxes are consistent within the statistical error, 
hence we used the result of region 'P'.
The best-fit spectra and parameters are shown in figure 6 and table 7, respectively.

\begin{table*}
  \caption{The best-fit parameters of M\,0.66$-$0.02}
  \label{tab:specfit_B2main2_free}
  \begin{center}
    \begin{tabular}{lccccccc}
      \hline\hline
      Observatory & Suzaku & \multicolumn{2}{c}{XMM-Newton} & \multicolumn{2}{c}{XMM-Newton} & Chandra & ASCA \\
      Detector & XIS(2005) & MOS(2004) & PN(2004) & MOS(2001) & PN(2001) & ACIS(2000) & SIS(1994) \\
      \hline
      \multicolumn{8}{c}{---Continuum---}\\
$N_{\rm H}$ & $5.0_{-0.3}^{+0.5}$ & $5.4_{-2.0}^{+2.1}$ & $3.7_{-1.1}^{+1.1}$ & $0.2_{-0.2}^{+2.4}$ & $2.5_{-2.5}^{+2.8}$ & $4.3_{-1.6}^{+1.1}$ & $3.0_{-1.4}^{+2.9}$ \\
$\Gamma$ & $3.0_{-0.3}^{+0.3}$ & $3.2_{-1.0}^{+0.5}$ & $1.7_{-0.3}^{+0.2}$ & $0.1_{-0.5}^{+1.0}$ & $1.2_{-1.1}^{+1.1}$ & $1.2_{-0.9}^{+0.5}$ & $1.7_{-1.4}^{+1.9}$ \\
$F_{\rm pow}$ & $2.0_{-0.1}^{+0.1}$ & $1.6_{-0.1}^{+0.1}$ & $2.4_{-0.1}^{+0.1}$ & $3.0_{-0.2}^{+0.2}$ & $2.4_{-0.2}^{+0.2}$ & $2.7_{-0.1}^{+0.1}$ & $2.3_{-0.2}^{+0.2}$ \\
      \multicolumn{8}{c}{---Neutral iron lines---}\\
$F_{640}$ & $0.66_{-0.02}^{+0.02}$ & $0.58_{-0.04}^{+0.04}$ & $0.57_{-0.04}^{+0.04}$ & $0.92_{-0.11}^{+0.11}$ & $0.82_{-0.09}^{+0.10}$ & $1.00_{-0.06}^{+0.05}$ & $0.99_{-0.14}^{+0.13}$ \\
$F_{706}$ & $0.06_{-0.04}^{+0.04}$ & $0.08_{-0.04}^{+0.04}$ & $0.02_{-0.02}^{+0.03}$ & $0.05_{-0.05}^{+0.09}$ & $0.11_{-0.07}^{+0.08}$ & $0.06_{-0.04}^{+0.06}$ & $0.12_{-0.11}^{+0.13}$ \\
      \multicolumn{8}{c}{---highly ionized iron lines---}\\
$F_{667}$ & $0.36_{-0.02}^{+0.02}$ & $0.29_{-0.04}^{+0.04}$ & $0.25_{-0.03}^{+0.03}$ & $0.42_{-0.09}^{+0.09}$ & $0.26_{-0.07}^{+0.08}$ & $0.25_{-0.04}^{+0.04}$ & $0.42_{-0.14}^{+0.13}$ \\
$F_{697}$ & $0.12_{-0.02}^{+0.03}$ & 0.10\footnotemark[$a$] & 0.08\footnotemark[$a$] & 0.14\footnotemark[$a$] & 0.09\footnotemark[$a$] & 0.08\footnotemark[$a$] & 0.14\footnotemark[$a$] \\
      \hline
$\delta_{\rm Gain}$ & 0(fix) & 0(fix) & $23_{-4}^{+6}$ & 0(fix) & $23_{-17}^{+13}$ & $-19_{-3}^{+8}$ & $81_{-11}^{+21}$ \\
$\chi^2$/dof & 348.4/354 & 36.9/40 & 71.3/68 & 35.3/38 & 16.6/27 & 142.1/144 & 36.1/37 \\
      \hline
      \multicolumn{8}{@{}l@{}}{\hbox to 0pt{\parbox{165mm}{\footnotesize
      	Note---Same as table 3.
      	\par\noindent
      	\footnotemark[$a$] The intensity of Fe \emissiontype{XXVI} K$\alpha$ is fixed at 34\% of that of Fe \emissiontype{XXV} K$\alpha$ obtained with Suzaku.
      }\hss}}
    \end{tabular}
  \end{center}
\end{table*}

\begin{figure}
  \begin{center}
    \FigureFile(80mm,50mm){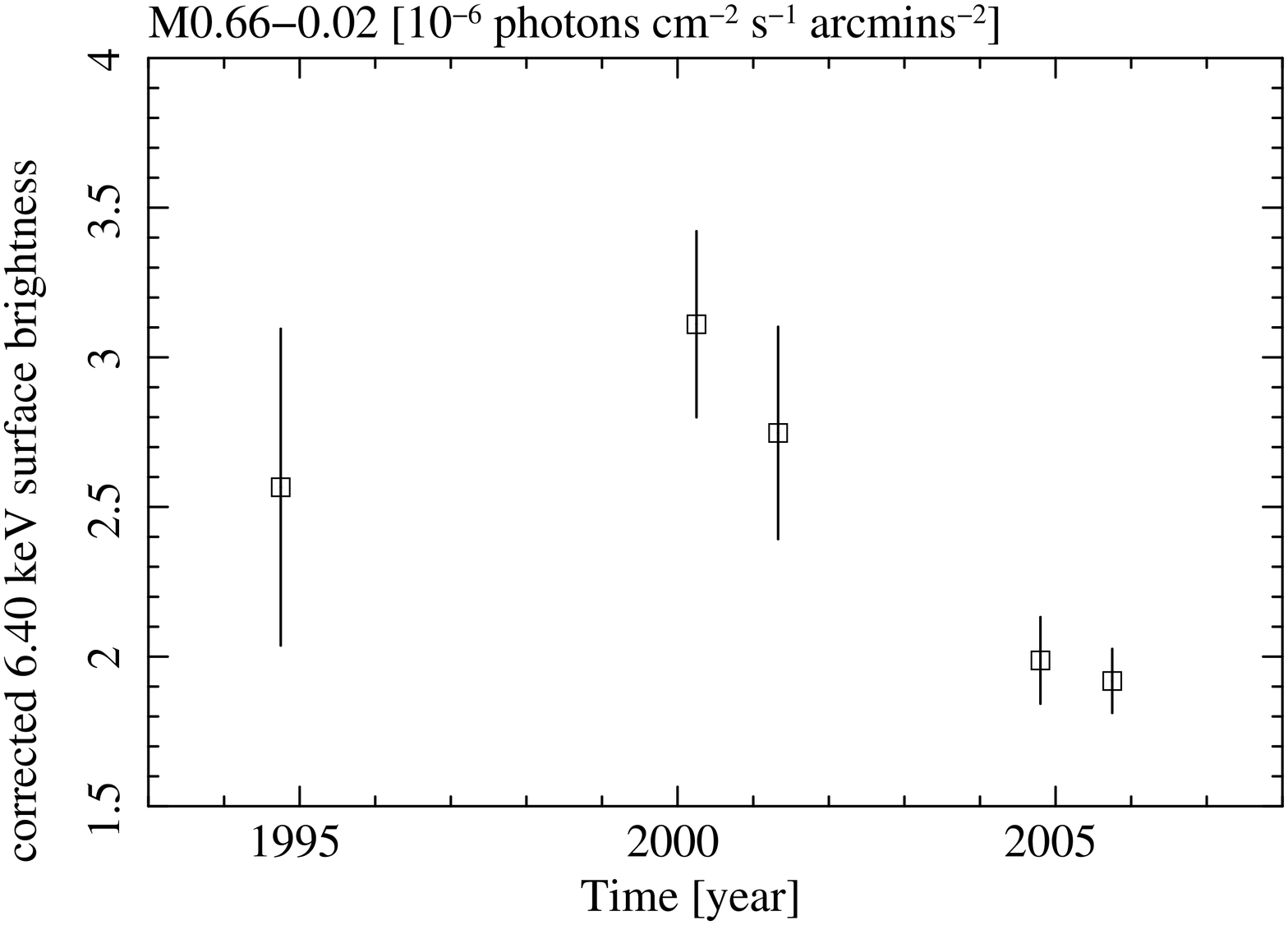}
  \end{center}
  \caption{The light curve of the 6.40~keV line surface brightness 
    from M\,0.66$-$0.02 (Sgr B2 cloud).
    In this plot, each error bar indicates a 90\% confidence limit.}
  \label{fig:trend_b2main_6.40_cor}
\end{figure}

However, cross-calibration errors for the efficiencies 
between the four satellites may not be neglected. 
We therefore corrected the 6.40~keV fluxes using the correction factors
shown in table \ref{tab:67normalizefactors}, 
which are derived from the 6.67~keV line flux in the Sgr B2 region 
(Section 3.5). 
The results are plotted in figure \ref{fig:trend_b2main_6.40_cor}.
This time history is similar to that of the Sgr B2 region 
(the whole region) (figure \ref{fig:trend_b2area_6.40_cor}).
The constant flux hypothesis is rejected with $\chi^2$/dof=30.7/6.
The flux in 2000 is 1.5 times higher than those in 2004 and 2005, 
and declines in 2001. 

\subsubsection{G\,0.570$-$0.018}

\begin{figure*}
  \begin{center}
    \FigureFile(50mm,50mm){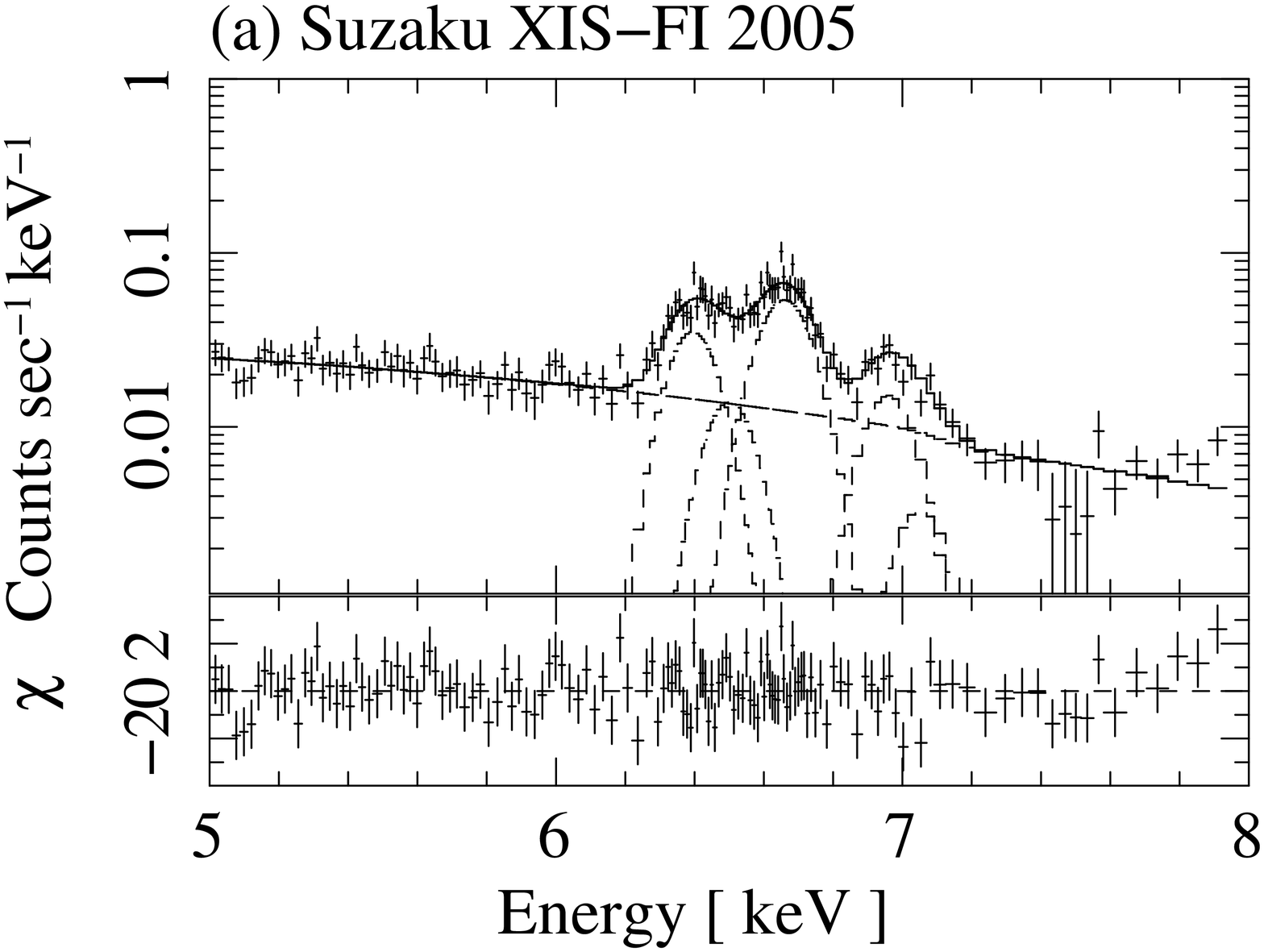}
    \FigureFile(50mm,50mm){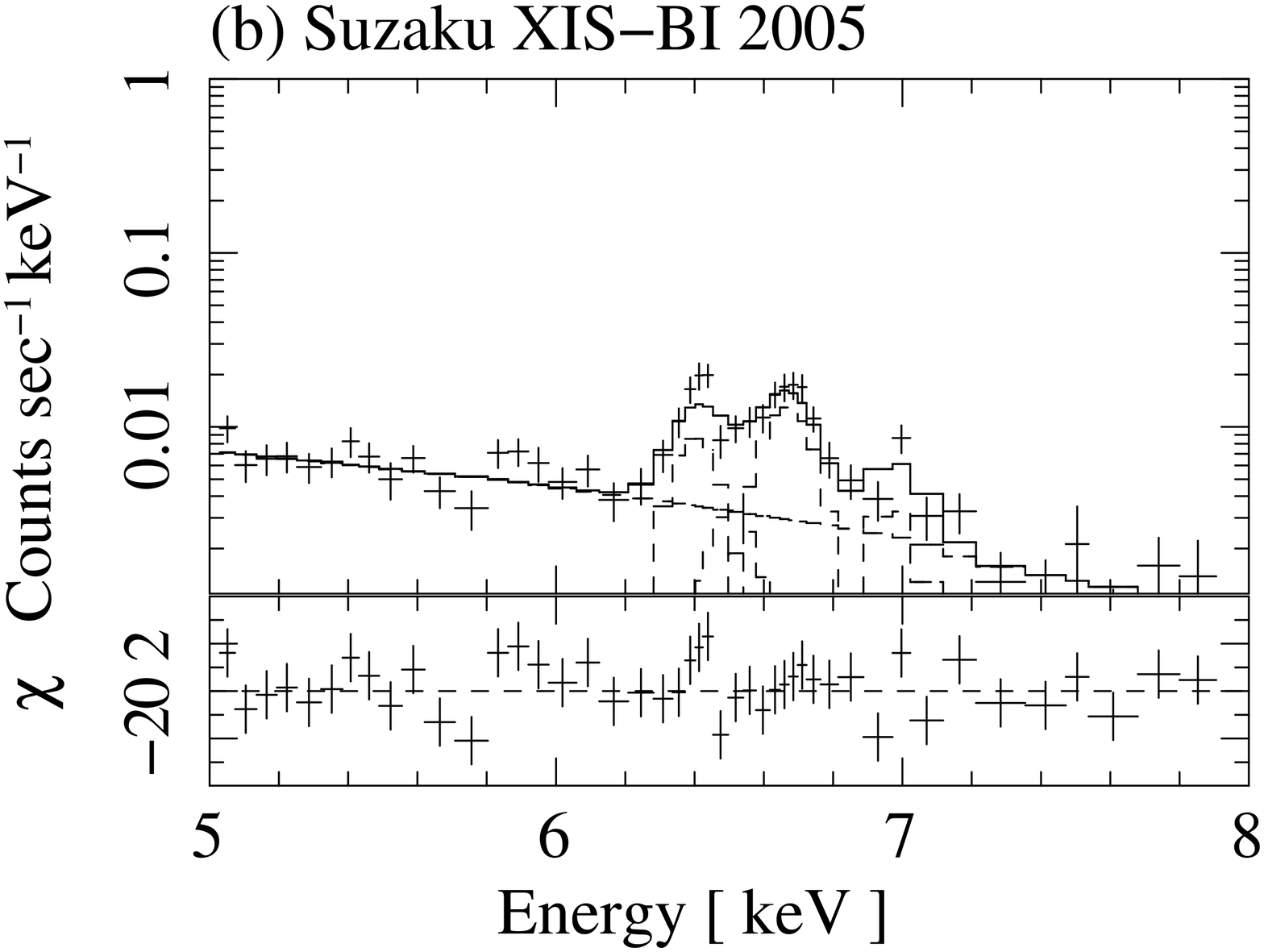}
    \FigureFile(50mm,50mm){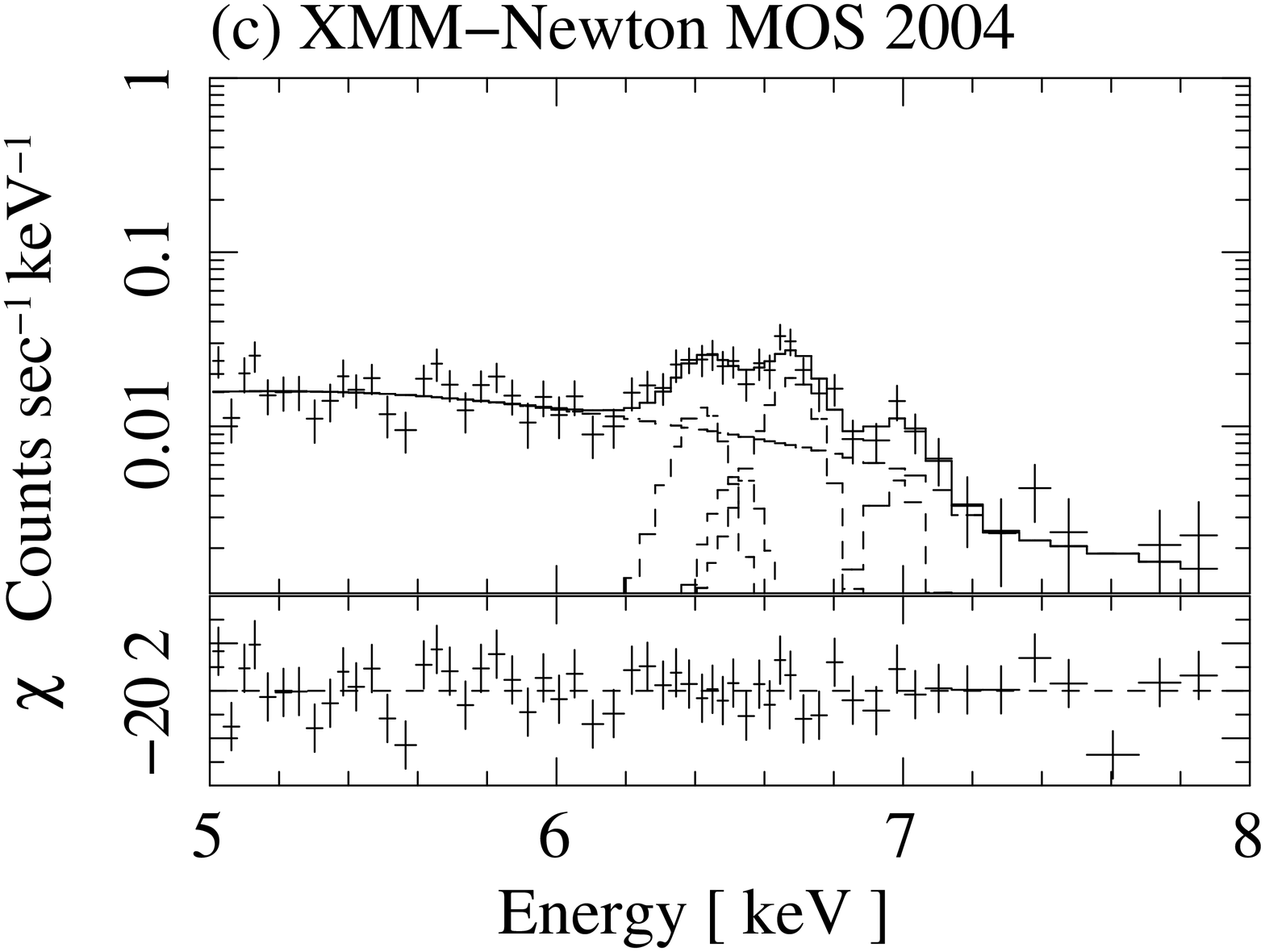}
    \FigureFile(50mm,50mm){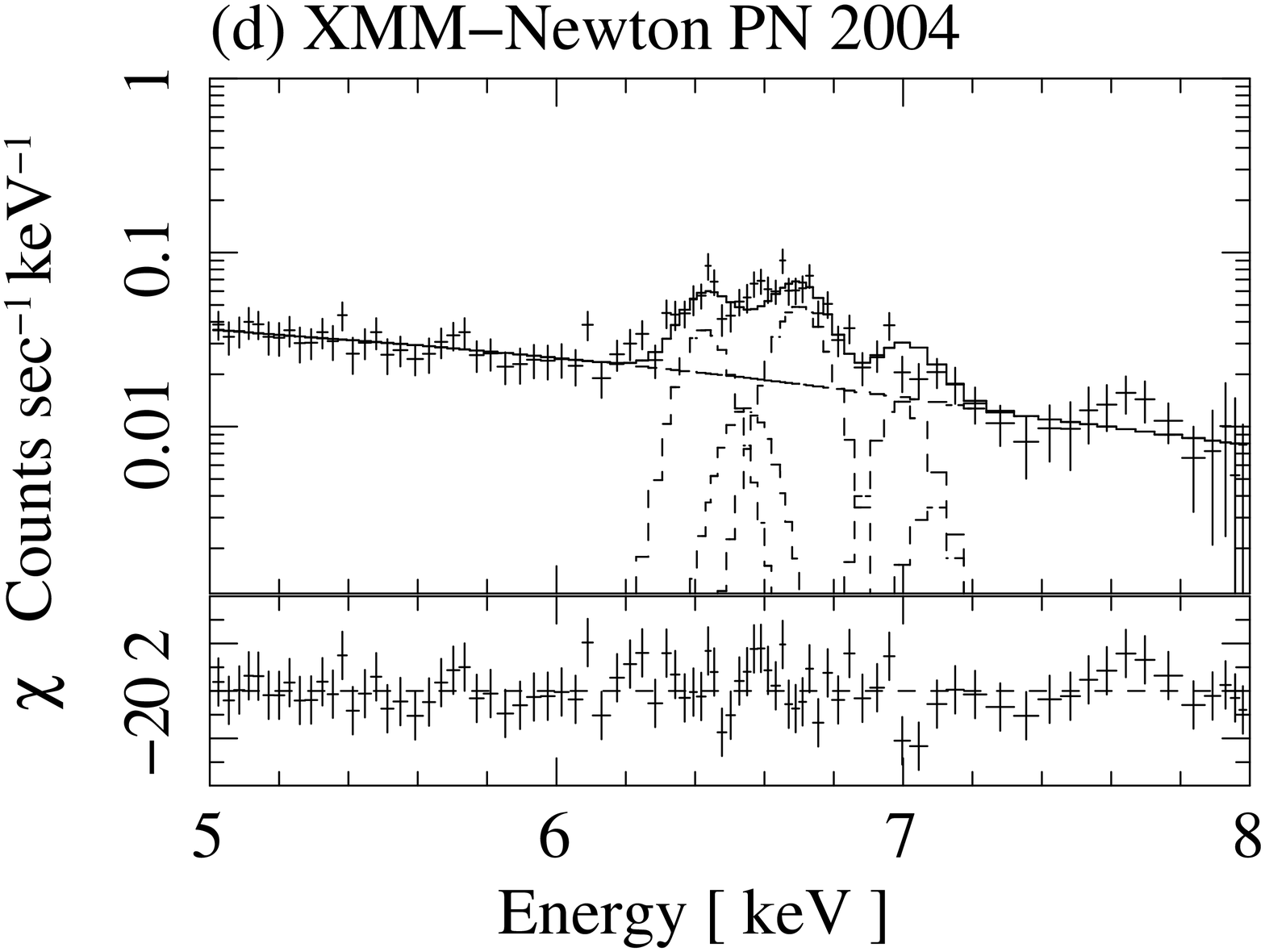}
    \FigureFile(50mm,50mm){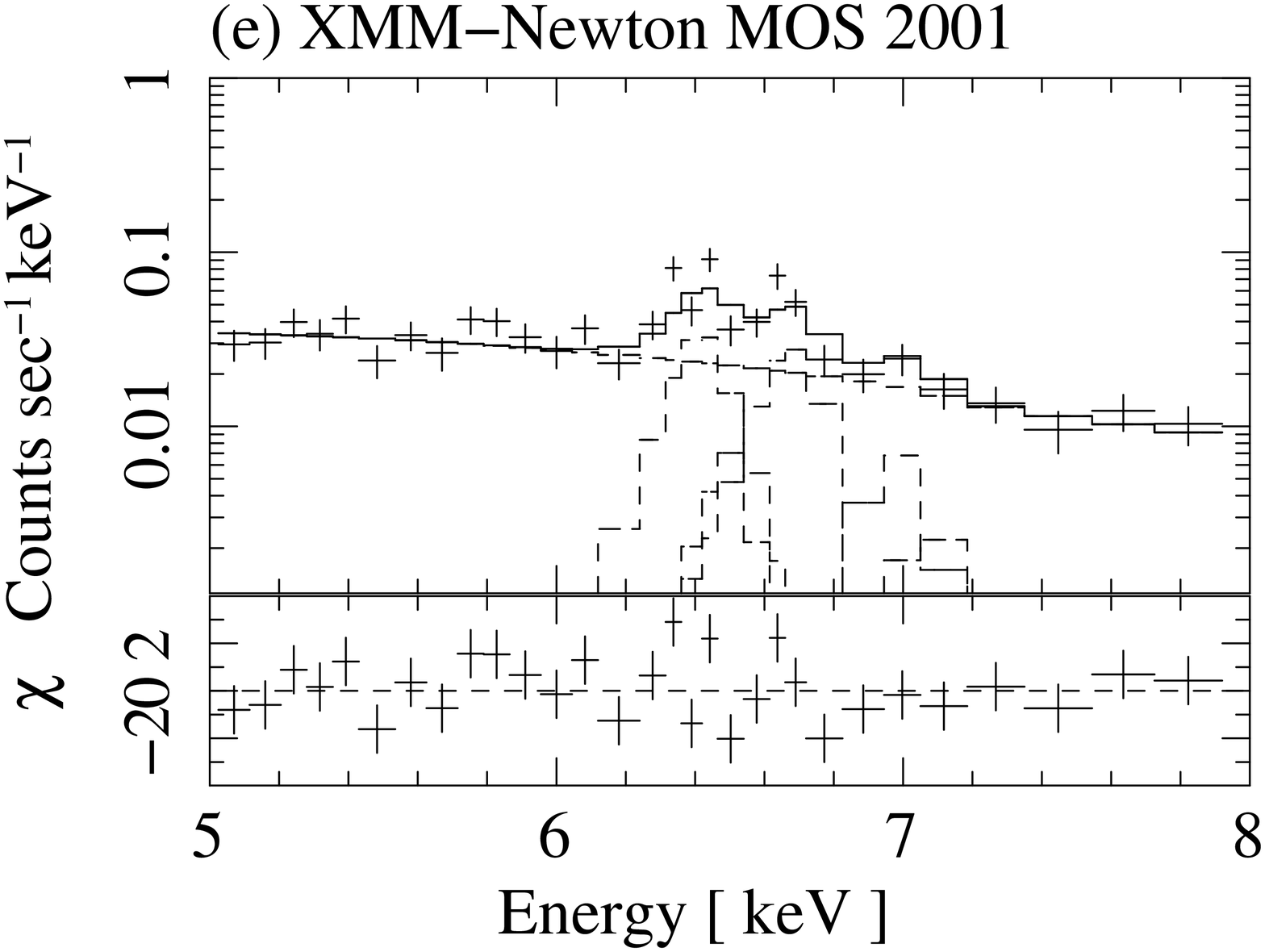}
    \FigureFile(50mm,50mm){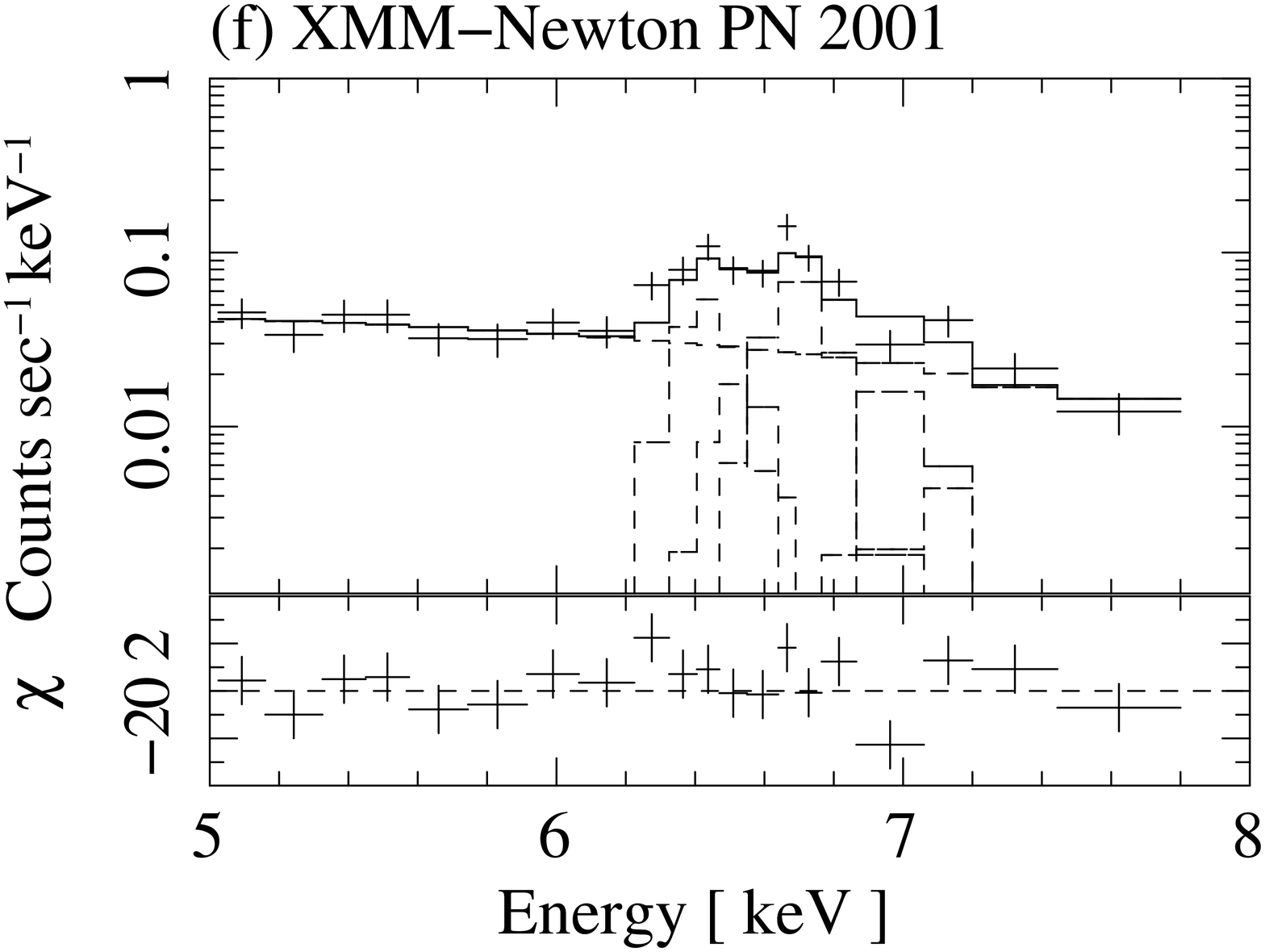}
    \FigureFile(50mm,50mm){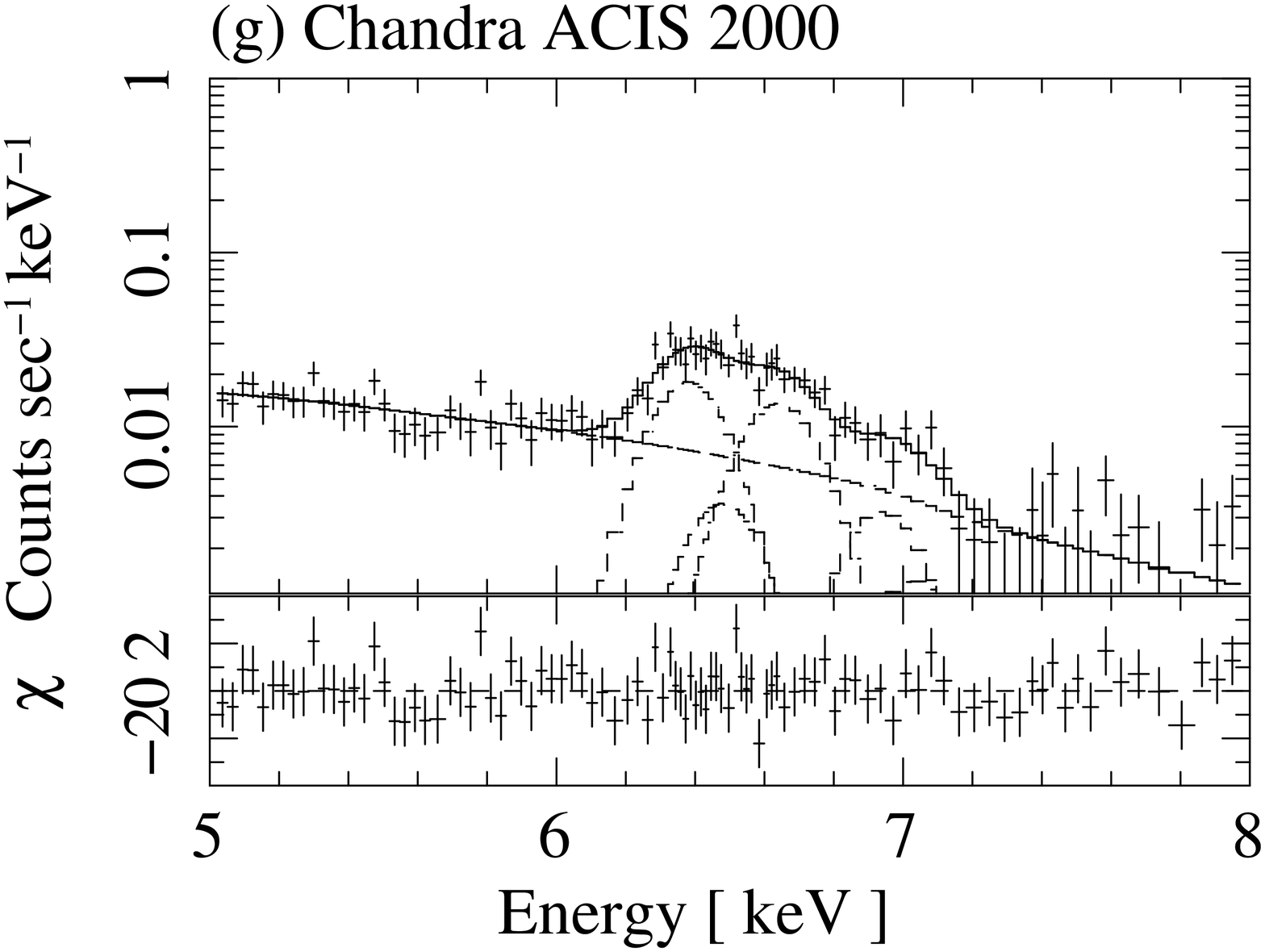}
    \FigureFile(50mm,50mm){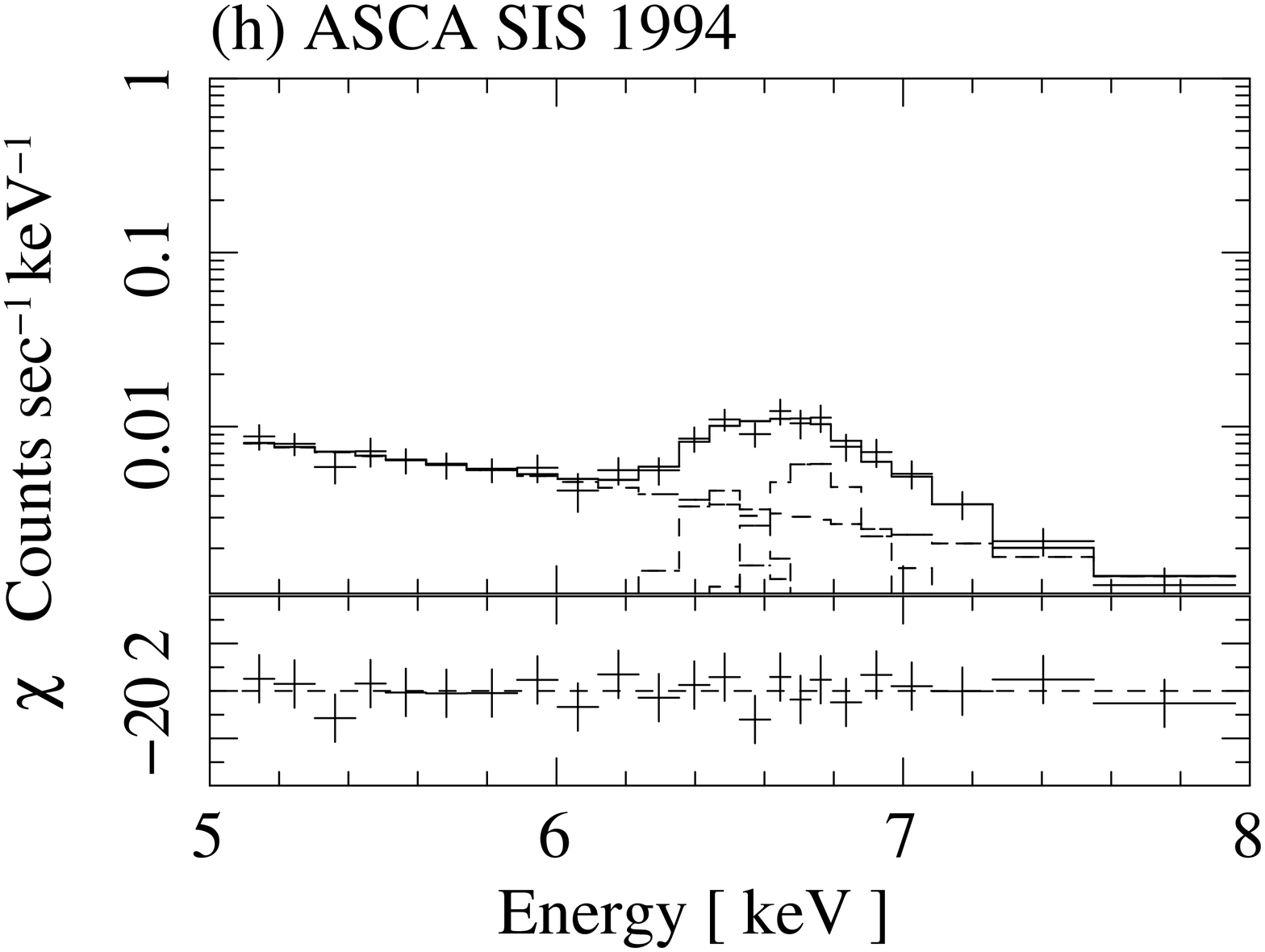}
  \end{center}
  \caption{Spectra for G\,0.570$-$0.018 obtained with Suzaku XIS FI sensors (a) and BI sensor (b) in 2005, 
    XMM-Newton MOS (c) and PN (d) in 2004, (e) and (f) in 2001, 
    Chandra ACIS-I (g) in 2000, and ASCA SIS (h) in 1994.
    The solid lines indicate the best-fit models.}
  \label{fig:spec_g0570}

\end{figure*}

G\,0.570$-$0.018 is a diffuse source 
extending to about 10 arcmin \citep{Senda2002}.
We extracted the 6.40~keV flux from G\,0.570$-$0.018 
(the dashed circle with a radius of 2.5~arcmin 
in figure \ref{fig:img_ironsb}) in the same way as M\,0.66$-$0.02.
Since the spectra do not have enough statistics 
to derive Fe \emissiontype{I} K$\beta$ (7.06~keV) line, 
the intensity of K$\beta$ was fixed to 12.5\% that of K$\alpha$ \citep{Kaastra1993}.
Using the same method as Section 3.5.1 except for the fixed line ratio, 
we fitted the Suzaku spectra with $N_{\rm H}$ and $\Gamma$ free model.
We obtained 6415~eV as the line centroid of the 6.40~keV line, which is
15~eV higher than the values obtained in the other region.
We then refitted the spectra with the line centroids fixed 
at the values for the Sgr B2 region (table \ref{tab:specfit_b2area_szk})
and found significant residuals in the 6.4--6.6~keV band.
We added one Gaussian function in this energy band, 
and obtained an acceptable fit.
The centroid of the additional line is 6505~eV, 
which is consistent with the value reported by \citet{Senda2002}.
From the energy centroid, this line is likely to be due to 
Fe$\emissiontype{XX}$-K$\alpha$.
Assuming a thin-thermal plasma, 
the temperature of the most prominent emission of this line is 1.1~keV,
a typical temperature of an SNR, 
which led \citet{Senda2002} to propose a new SNR 
in highly non-equilibrium ionization.

\begin{table*}
  \caption{The best-fit parameters of G\,0.570$-$0.018}
  \label{tab:specfit_g0570_free}
  \begin{center}
    \begin{tabular}{lccccccc}
      \hline\hline
      Observatory & Suzaku & \multicolumn{2}{c}{XMM-Newton} & \multicolumn{2}{c}{XMM-Newton} & Chandra & ASCA \\
      Detector & XIS(2005) & MOS(2004) & PN(2004) & MOS(2001) & PN(2001) & ACIS(2000) & SIS(1994) \\
      \hline
 \multicolumn{8}{c}{---Continnum---}\\
      $N_{\rm H}$ & $1.5_{-1.0}^{+1.3}$ & $6.2_{-2.5}^{+1.6}$ & $0.4_{-0.4}^{+1.7}$ & $1.3_{-1.3}^{+1.4}$ & $1.5_{-1.5}^{+3.6}$ & $2.0_{-2.0}^{+2.7}$ & $0.2_{-0.2}^{+3.4}$ \\
      $\Gamma$ & $2.6_{-0.5}^{+0.6}$ & $4.1_{-1.6}^{+1.9}$ & $1.8_{-0.6}^{+0.8}$ & $0.5_{-0.9}^{+1.2}$ & $2.0_{-1.2}^{+1.6}$ & $2.0_{-0.6}^{+1.4}$ & $1.2_{-0.8}^{+1.5}$ \\
      $F_{\rm pow}$ & $1.1_{-0.1}^{+0.1}$ & $1.1_{-0.1}^{+0.1}$ & $1.3_{-0.1}^{+0.1}$ & $1.9_{-0.1}^{+0.1}$ & $1.2_{-0.1}^{+0.1}$ & $1.2_{-0.1}^{+0.1}$ & $0.9_{-0.1}^{+0.1}$ \\
      \multicolumn{8}{c}{---Neutral iron lines---}\\
$F_{640}$ & $0.16_{-0.05}^{+0.03}$ & $0.11_{-0.04}^{+0.04}$ & $0.14_{-0.03}^{+0.03}$ & $0.18_{-0.06}^{+0.06}$ & $0.19_{-0.06}^{+0.06}$ & $0.26_{-0.04}^{+0.04}$ & $0.12_{-0.05}^{+0.05}$ \\
      \multicolumn{8}{c}{---Ionized iron lines---}\\
$F_{667}$ & $0.29_{-0.02}^{+0.02}$ & $0.20_{-0.04}^{+0.04}$ & $0.22_{-0.03}^{+0.03}$ & $0.18_{-0.06}^{+0.06}$ & $0.23_{-0.06}^{+0.06}$ & $0.25_{-0.03}^{+0.03}$ & $0.20_{-0.02}^{+0.04}$ \\
$F_{651}$ & $0.07_{-0.02}^{+0.03}$ & 0.05\footnotemark[$d$] & 0.06\footnotemark[$d$] & 0.04\footnotemark[$d$] & 0.05\footnotemark[$d$] & 0.06\footnotemark[$d$] & 0.05\footnotemark[$d$] \\
      \hline
      $\delta_{\rm Gain}$ & 0(fix) & 0(fix) & 35(fix) & 0(fix) & 33(fix) & $-18$(fix) & 71(fix) \\
      $\chi^2$/dof & 192.3/181 & 56.1/51 & 70.8/78 & 40.9/25 & 21.0/15 & 82.4/93 & 6.8/18 \\
      \hline
      \multicolumn{8}{@{}l@{}}{\hbox to 0pt{\parbox{165mm}{\footnotesize
      	Note---Same as table 3.
	\par\noindent
      	\footnotemark[$d$] Fixed at the value and flux ratio against Fe \emissiontype{XXV} K$\alpha$ line obtained with Suzaku.
      }\hss}}
    \end{tabular}
  \end{center}
\end{table*}

Since the flux of the 6.51~keV line is due to highly ionized iron 
and the flux is likely to be time constant, 
the flux ratio among the highly ionized iron lines 
is fixed in the following analysis. 
The best-fit spectra and parameters are shown in figure 8 and table 8, respectively.

\begin{figure}
  \begin{center}
    \FigureFile(80mm,50mm){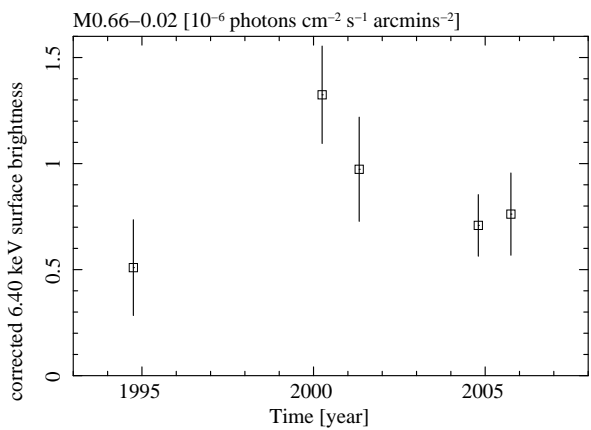}
  \end{center}
  \caption{The light curve of the 6.40~keV line surface brightness 
    from G\,0.570$-$0.018. 
    In this plot, each error bar indicates a 90\% confidence limit.}
  \label{fig:trend_g0570_6.40_cor}
\end{figure}

We corrected the 6.40~keV fluxes using the correction factors 
as we did for M\,0.66$-$0.02.
Figure \ref{fig:trend_g0570_6.40_cor} shows the time trend 
of the 6.40~keV line from G\,0.570$-$0.018, 
which indicates a clear variability in the 6.40~keV line emission. 
A constant flux hypothesis is marginally rejected with $\chi^2$/dof = 13.2/6.

\citet{Senda2002} mentioned that G\,0.570$-$0.018 is 
a simple SNR candidate, but the 6.40~keV variability suggests that 
G\,0.570$-$0.018 includes an XRN (see Discussion). 
In fact, \citet{Miyazaki2000} reported a flux peak 
of the CS ($J$=1-0) line emission at 
$(l,~b) = (\timeform{0D.601},~-\timeform{0D.025})$, 
which indicates the presence of a molecular cloud.
The light curve of G\,0.570$-$0.018 is similar to 
that of M\,0.66$-$0.02. 
Hence the distances of the two clouds from Sgr A$^*$ are the same. 
G\,0.570$-$0.018 may be located in front of M\,0.66$-$0.02 
with some off-sets of the line of sight.

\section{Discussion}

The 6.40~keV line is K-shell emission of neutral 
or low-ionization iron. Hence, since cool gases such as molecular clouds 
cannot emit X-ray photons, 
the 6.40~keV line is due to illumination by external sources 
of either X-rays or charged particles.
The analyses of ASCA, Chandra and Suzaku 
(\cite{Koyama1996}, \cite{Murakami2001sgrb}, \cite{Koyama2007b}) 
revealed strong Fe\emissiontype{I} K$\alpha,\beta$ lines 
and deep Fe K edge, which supports an X-ray origin.
However, an electron origin cannot be excluded 
if the iron abundance is higher than the solar value 
by a factor of 3--4 \citep{Yusef2007} 
and the X-ray spot is at the back-side of the molecular clouds.

The time variability of the 6.40~keV line in the Sgr B2 region 
has already been reported by \citet{Koyama2008a} 
from the Chandra (2000) to the Suzaku (2005) observations.
A similar time variability has also been reported 
for the other cloud near Sgr A$^*$ by \citet{Muno2007} 
and \citet{Koyama2008b}, 
which are the time variabilities of a few years span. 
In this paper, we have confirmed the time variability 
with a longer time scale ($\sim$10 years) from the Sgr B2 region. 

This time variability is also found in the sub-structures, 
M\,0.66$-$0.02 and G\,0.570$-$0.018. 
The sizes of these two sources are 
a few tens of light years, which is comparable to the variation time scale.
These time variabilities are the clearest evidence 
for X-ray origin, because it is difficult to change the flux of any charged particle in a large region (a few~10 light-years) 
within a few~10 years.

If the origin is X-ray irradiation, 
the observed 6.40~keV flux requires a bright source 
with a luminosity of more than $10^{37}$ erg s$^{-1}$ 
assuming the source is located near the Sgr B2 region 
in the distance of 8.5~kpc.
Point sources in the GC region have been catalogued 
by many satellites, but no source 
has been brighter than $10^{37}$ erg s$^{-1}$ 
in a time scale of ten years. 
Thus, we conclude that the most probable source is 
the massive black hole Sgr A$^*$.

In this scenario, the light curves of the 6.40~keV line flux 
in the molecular clouds 
can be used as a tracer of the light curve of Sgr A$^*$ in the past, 
like a time-delayed echo.
Since the light travel time between Sgr A$^*$ and the Sgr B2 region 
is $\sim$300~years, 
we already proposed that Sgr A$^*$ was X-ray bright 
($\sim$ 10$^{39}$ erg s$^{-1}$) 300 years ago. 
Our new results indicate that the X-rays have been variable 
over a time scale of a few tens of years.
The light curve of the 6.40 keV line is not the real history of Sgr A$^*$.
It is smeared out if the size of the reflection (fluorescence) cloud is comparable 
or larger than the light-cross scale.
Then, (1) even if the real light curve is pulse-like (delta-function), 
the observed light curve of the fluorescent line is smoothed on a scale of 
the light crossing time of the cloud, or 
(2) if the cloud is very small (like a point source), the observed light 
curve is the real history of the incident X-rays. 
The high-resolution Chandra image \citep{Koyama2008a} revealed that 
the fluorescent line emission has a complex structure with bright peak of 
a scale of less than 1 arcmin ($\sim$6 light years). 
Therefore, a realistic situation lies between cases (1) and (2), 
so that the observed light curve is moderately smeared-out. 
This means that the observed variability amplitude is a lower limit of 
the actual one, although they do not differ largely from each other. 
Hereafter, we discuss the evolution of the Sgr A$^*$ outburst 
based on the observed light curve.
The averaged decay rate of the 6.40~keV line is 60\% in 10 years, 
hence the half-decay time is $\sim$15~years. 
If this decay time has been constantly folded for 300 years, 
the X-ray luminosity after 300 years becomes 
10$^{33\mbox{--}34}$ erg s$^{-1}$, which is 
consistent with the observed value in the present. 
We hence propose that 
300 years ago, Sgr A$^*$ experienced a giant flare 
and then entered a decay phase that continued to the present.

\section{Summary}
We summarize the results as follows:
\begin{enumerate}
	
\item We have extensively studied the spectra of the Sgr B2 region, assuming various continuum models, 
  backgrounds and uncertainties of the relative efficiencies of Suzaku, XMM-Newton Chandra, and ASCA. 
  We then found that the 6.40~keV line fluxes are time variable.

\item We found the time variability of the individual clouds, M\,0.66$-$0.02 and G\,0.570$-$0.018.

\item The time histories of the 6.40~keV line flux are similar for all the sources, 
  and exhibit their brightest peak in 2000, decline in 2001 and finally fall to 60\% in 2005.

\item The 6.40~keV variability of the diffuse source G\,0.570$-$0.018 suggests that this source is a candidate for an XRN.

\item  The variability of the 6.40~keV line indicates that 
  the GC black hole Sgr A$^*$ experienced a large flare 300 years ago and entered a decay phase. 

\end{enumerate}

\bigskip
The authors thank all of the Suzaku team  members, especially H. Uchiyama, H. Nakajima, H. Yamaguchi, 
and H. Mori for their support and for providing useful information on the XIS performance.
This work is supported by Grant-in-Aids from the Ministry of Education, Culture, Sports, Science and Technology (MEXT) of Japan, 
the 21st Century COE "Center for Diversity and Universality in Physics'', Scientific Research A (KK), 
Priority Research Areas in Japan ``New Development in Black Hole Astronomy''(TGT), and Grant-in-Aid for Young Scientists B (HM) 
HM is also supported by the Sumitomo Foundation, Grant for Basic Science Research Projects, 071251, 2007. 
TI is supported by a JSPS Research Fellowship for Young Scientists.


\begin{thebibliography}{99}

\bibitem[Baganoff \etal (2001)]{Baganoff2001} Baganoff, F.~K., et 
al.\ 2001, \nat, 413, 45 

\bibitem[Carter \& Read (2007)]{Carter2007} Carter, J.~A., \& Read, 
A.~M.\ 2007, \aap, 464, 1155 

\bibitem[Eisenhauer et al.(2005)]{Eisenhauer2005} Eisenhauer, F., et 
al.\ 2005, \apj, 628, 246

\bibitem[Ghez et al.(2005)]{Ghez2005} Ghez, A.~M., Salim, S., 
Hornstein, S.~D., Tanner, A., Lu, J.~R., Morris, M., Becklin, E.~E., \& 
Duch{\^e}ne, G.\ 2005, \apj, 620, 744 

\bibitem[Kaastra \& Mewe (1993)]{Kaastra1993} Kaastra, J.~S., \& 
Mewe, R.\ 1993, \aaps, 97, 443 

\bibitem[Koyama \etal (1996)]{Koyama1996} Koyama, K., Maeda, Y., Sonobe, T., Takeshima, T., Tanaka, Y., \& Yamauchi, S.\ 1996, \pasj, 48, 249 

\bibitem[Koyama \etal (2007a)]{Koyama2007a} Koyama, K. et al. \ 2007a, \pasj, 59, S23

\bibitem[Koyama \etal (2007b)]{Koyama2007b} Koyama, K. et al. \ 2007b, \pasj, 59, S221 

\bibitem[Koyama \etal (2007c)]{Koyama2007c} Koyama, K. et al. \ 2007c, \pasj, 59, S245 

\bibitem[Koyama \etal (2008a)]{Koyama2008a} Koyama, K. et al. \ 2008a, \pasj, 60, (in press) (astroph/0711.2853) 

\bibitem[Koyama \etal (2008b)]{Koyama2008b} Koyama, K. et al. \ 2008b, \pasj, submitted (\#3307) 

\bibitem[Mitsuda \etal (2007)]{Mitsuda2007} Mitsuda, K. et al. \ 2007, \pasj, 59, S1

\bibitem[Miyazaki \& Tsuboi (2000)]{Miyazaki2000} Miyazaki, A., \& Tsuboi, M.\ 2000, \apj, 536, 357 

\bibitem[Muno \etal (2003)]{Muno2003} Muno, M.~P., et al.\ 2003, \apj, 589, 225 
\bibitem[Muno \etal (2007)]{Muno2007} Muno, M.~P., Baganoff, 
F.~K., Brandt, W.~N., Park, S., \& Morris, M.~R.\ 2007, \apjl, 656, L69

\bibitem[Murakami et al.(2001a)]{Murakami2001sgrc} Murakami, H., Koyama, 
K., Tsujimoto, M., Maeda, Y., \& Sakano, M.\ 2001, \apj, 550, 297 

\bibitem[Murakami \etal (2001b)]{Murakami2001sgrb} Murakami, H., Koyama, K., \& Maeda, Y.\ 2001, \apj, 558, 687 

\bibitem[Nobukawa et al. (2008)]{Nobukawa2008} Nobukawa, M., et al. 2008, \pasj, 60, (in press) (astroph/0712.0877) 

\bibitem[Predehl et al. (2003)]{Predehl2003} Predehl, P., Costantini, E., Hasinger, G., \& Tanaka, Y.  2003, Astronomische Nachrichten, 324, 73

\bibitem[Reid \etal (1988)]{Reid1988} Reid, M.~J., Schneps, M.~H., Moran, J.~M., Gwinn, C.~R., Genzel, R., Downes, D., \& Roennaeng, B.\ 1988, \apj, 330, 809 

\bibitem[Schwartz \etal (2000)]{Schwartz2000} Schwartz, D.~A., et 
al.\ 2000, \procspie, 4012, 28

\bibitem[Senda \etal (2002)]{Senda2002} Senda, A., Murakami, H., \& Koyama, K.\ 2002, \apj, 565, 1017 

\bibitem[Serlemitosos \etal (2007)]{Serlemitosos2007} Serlemitsos, P. et al. \ 2007, \pasj, 59, S9 

\bibitem[Str{\"u}der \etal (2001)]{Struder2001} Str{\"u}der, L., et 
al.\ 2001, \aap, 365, L18 

\bibitem[Turner \etal (2001)]{Turner2001} Turner, M.~J.~L., et 
al.\ 2001, \aap, 365, L27

\bibitem[Tatischeff (2003)]{Tatischeff2003} Tatischeff, V. 2003, in Final Stage of Stellar Evolution, ed. C. Motch \& Hameury (EAS publication Series vol.7), 79 (astro-ph/0208397v1)

\bibitem[Weisskopf \etal (2002)]{Weisskopf2002} Weisskopf, M.~C., 
Brinkman, B., Canizares, C., Garmire, G., Murray, S., \& Van Speybroeck, 
L.~P.\ 2002, \pasp, 114, 1 

\bibitem[Weisskopf \etal (2003)]{Weisskopf2003} Weisskopf, M.~C., et 
al.\ 2003, Experimental Astronomy, 16, 1

\bibitem[Yusef-Zadeh \etal (2007)]{Yusef2007} Yusef-Zadeh, F., 
Muno, M., Wardle, M., \& Lis, D.~C.\ 2007, \apj, 656, 847

\end{thebibliography}
\end{document}